\documentclass{article}
\usepackage{latexsym}
\usepackage{graphicx, epsfig}
\usepackage{amsmath, amsthm}

\newenvironment{junk}[1]{}

\title{A bounded-degree network formation game}

\author{\begin{tabular}[t]{ccc}
Nikolaos Laoutaris & Rajmohan Rajaraman & Ravi Sundaram\\[0.5
mm] Harvard University & Northeastern University & Northeastern University\\[0.5
mm] \tt{nlaout@eecs.harvard.edu} & \tt{rraj@ccs.neu.edu} &
\tt{koods@ccs.neu.edu} \\ [20 pt] & Shang-Hua Teng & \\[0.5
mm] & Boston University & \\[0.5
mm] & \tt{steng@cs.bu.edu} &
\end{tabular}}

\begin{document}
\date{}

\newtheorem{definition}{Definition}
\newtheorem{lemma}{Lemma}
\newtheorem{theorem}{Theorem}
\newtheorem{conjecture}{Conjecture}
\newtheorem{corollary}{Corollary}

\maketitle

\begin{abstract}
Motivated by applications in peer-to-peer and overlay networks we
define and study the \emph{Bounded Degree Network Formation}
(BDNF) game. In an $(n,k)$-BDNF game, we are given $n$ nodes, a
bound $k$ on the out-degree of each node, and a weight $w_{vu}$
for each ordered pair $(v,u)$ representing the traffic rate from
node $v$ to node $u$. Each node $v$ uses up to $k$ directed links
to connect to other nodes with an objective to minimize its
average distance, using weights $w_{vu}$, to all other
destinations. We study the existence of pure Nash equilibria for
$(n,k)$-BDNF games. We show that if the weights are arbitrary,
then a pure Nash wiring may not exist. Furthermore, it is NP-hard
to determine whether a pure Nash wiring exists for a given
$(n,k)$-BDNF instance. A major focus of this paper is on uniform
$(n,k)$-BDNF games, in which all weights are 1. We describe how to
construct a pure Nash equilibrium wiring given any $n$ and $k$,
and establish that in all pure Nash wirings the cost of individual
nodes cannot differ by more than a factor of nearly 2, whereas the
diameter cannot exceed $O(\sqrt{n \log_k n})$. We also analyze
best-response walks on the configuration space defined by the
uniform game, and show that starting from any initial
configuration, strong connectivity is reached within $\Theta(n^2)$
rounds. Convergence to a pure Nash equilibrium, however, is not
guaranteed. We present simulation results that suggest that
loop-free best-response walks always exist, but may not be
polynomially bounded. We also study a special family of
\emph{regular} wirings, the class of Abelian Cayley graphs, in
which all nodes imitate the same wiring pattern, and show that if
$n$ is sufficiently large no such regular wiring can be a pure
Nash equilibrium.
\end{abstract}
\section{Introduction}
\label{sec:intro}
In this paper, we define and study a graph-theoretic game, called the
\emph{Bounded Degree Network Formation} (BDNF) game. An $(n,k)$-BDNF
game models the distributed formation of a directed network with $n$
nodes in which each node selfishly selects $k$ out-going links, with
the goal of minimizing a cost that is a function of its distances to
other nodes in the network.  We focus on the following cost function:
For each ordered pair $(v,u)$ of nodes, we have a weight $w_{vu}$ that
captures the traffic rate from $v$ to $u$, and we define the {\em
cost}\/ of a node $v$ as the total weighted distance to all the other
nodes.  In this setting, we call a network or a wiring {\em stable} if
no node can improve its cost by an unilateral rewiring. In other
words, a stable wiring represents a pure Nash equilibrium of the BDNF
game.

Bounded-degree network formation games arise in the context of popular
applications like \emph{unstructured peer-to-peer (P2P)
file-sharing}~\cite{Lv02:Gia,Chawathe03:GnutellaScalable} and
\emph{overlay routing}~\cite{Andersen01:RON} when the
participating nodes behave strategically and select first-hop
neighbors in order to selfishly optimize their own utility. In both
applications, bounded degrees are in place --- albeit for different
reasons. In unstructured P2P file-sharing, nodes employ
scoped-flooding or multiple parallel random walks to reach other nodes
and thus have to adhere to small out-degrees in order to protect the
network from getting clogged with queries. Overlay routing systems
have been proposed for allowing nodes to route their traffic through
alternative overlay paths that offer better resilience or quality of
service than the standard ones offered by the native IP routing
mechanism. Such systems run full fledged link-state routing protocols
at the overlay layer and thus have to monitor and disseminate
link-state information for all the overlay links. Initial systems,
like RON~\cite{Andersen01:RON}, assumed a full-mesh overlay topology
and as a consequence did not scale well (could go up to around 50
nodes). Subsequent proposals~\cite{Young04:KMST,Liu05:OverlayUnderlay}
have achieved better scalability by putting constraints on the
out-degree of nodes so as to reduce the number of overlay links that
need to be monitored.

The second defining characteristic of our network formation game --
the directionality of links -- is also well-reflected in several
applications from both families. In P2P file-sharing, link
directionality is usually explicitly specified in the protocols used
for implementing the system, whereas in routing, it arises as a
consequence of the employed business
strategies~\cite{Anshelevich06:FOCS}.  Considering the aspect of
routing (for queries or traffic), we note that our choice of the cost
function for the BDNF game implicitly assumes that shortest-path
routing is in place.  This is indeed the case in overlay routing
systems, where the nodes have global awareness of the overlay topology
by participating in the link-state overlay routing protocol, and thus
can execute a shortest-path algorithm and determine such routes. In
P2P file-sharing systems that employ scoped flooding or random walks,
routing can deviate from being shortest-path.  We notice, however,
that it can be made approximately close to being shortest-path by
increasing the scope of flooding or the density of parallel random
walks (at an extreme case, full flooding guarantees that a destination
will be reached through a shortest-path).

\subsection{Our Results}
In this paper, we study the structural and complexity-theoretic
properties of stable wirings.  We first consider general non-uniform
games, in which the weights are arbitrary.
\begin{itemize}
\item
For any $k$, and any $n$ sufficiently large, there exists a
collection of weights $w_{vu}$ for which the $(n,k)$-BDNF game has
no pure Nash equilibrium.  Furthermore, it is NP-hard to determine
whether a pure Nash wiring exists for a given $(n,k)$-BDNF
instance with arbitrary weights.  These results are in
Section~\ref{sec:nonuniform}.
\end{itemize}
The main focus of this paper is on uniform games, where all the
weights are 1.
\begin{itemize}
\item
One of our main results is a proof that every uniform $(n,k)$-BDNF
game has a pure Nash equilibrium wiring.  Our proof is constructive
and our stable wiring enjoys the property that the radius of each node
is at most $2\log_k n -1$, implying that the total distance is at most
twice as much as the best possible network that could be constructed by a
central network designer.  This result appears in
Section~\ref{sec:uniform.exist}
\item
Although the complete characterization of stable wirings for
uniform $(n,k)$-BDNF games remains an open research problem, we
establish some general properties of all stable wirings.  We show
that every stable wiring for a uniform $(n,k)$-uniform game is
almost fair: the cost of any node is within $2 + 1/k + o(1)$ times
the cost of any other node.  We also explore the possibility of
completely fair stable wirings by studying regular wiring patterns
formed by Abelian Cayley graphs, a special class of
vertex-transitive graphs.  We show that for any $k \geq 2$, no
Abelian Cayley graph with degree $k$ and $n$ nodes is a pure Nash
equilibrium for $n > c2^k$ for some constant $c$. These results
appear in Section~\ref{sec:uniform.structure}.
\end{itemize}
\junk{Some
non Abelian Cayley vertex transitive graphs, such as the directed
version of the Peterson graph, are also not able. In contrast, for $k
\geq n/2$, stable Abelian Cayley graphs exist.}
Our final set of results, which appear in Section~\ref{sec:best},
concern best-response walks on the configuration space defined by
uniform games.
\begin{itemize}
\item
We show that starting from any initial configuration, strong
connectivity is reached within $\Theta(n^2)$ steps.  Convergence to a
pure Nash equilibrium, however, is not guaranteed.  We present
simulation results that suggest that loop-free best-response walks
always exist, but may not be polynomially bounded.
\end{itemize}
Our experiments have suggested several interesting open questions and
conjectures, which we present in Section~\ref{sec:open}.

\junk{
\label{subsec:results} We define a peering game as follows:
there are $n$ nodes; each node has a budget of outgoing links it
can use to connect to other nodes; each node uses its budget to
minimize its average distance to the rest of the nodes. The
equal-budget case refers to the situation where each node has the
same budget. $k$-budget refers to the situation where each node
has the same budget of $k$ outgoing links. The uniform case refers
to the situation where the average distance is taken uniformly
over all other nodes while in the nonuniform case the distance to
different nodes can be weighted differently. Our main results are
as follows:
\begin{itemize}
    \item For all $n$ there exist $1$-budget $n$-node nonuniform peering games
    with no pure Nash equilibria; in fact it is NP-hard to determine
    whether there exists a pure Nash Equilibrium or not.
    \item For every $k$ and $n$ the $k$-budget $n$-node
    uniform peering games has a pure Nash equilibrium.
\end{itemize}
We also prove a number of smaller results that we will present
after introducing some requisite terminology and definitions. }

\subsection{Related Work}
\label{sec:related}
Our model for network formation is inspired in large part by
\cite{fabrikant03network} where they defined and studied a similar
network creation game.  Rather than assuming a fixed budget of
outgoing links as in our network formation game the authors
in~\cite{fabrikant03network} consider undirected links, and the nodes
optimize a cost which is the sum of the number of edges, scaled by a
parameter $\alpha > 0$, and the sum of distances to the rest of the
nodes.  They present several results on the price of anarchy, which is
the ratio of the cost of the worst-case Nash equilibrium to the social
optimum cost~\cite{koutsoupias99worst}.  Further results on this
network formation are obtained in~\cite{albers+eemr:nash}.
In~\cite{moscibroda06topologies} a variant of this
game is studied, in which the nodes are embedded in a metric space and
the distance component of the cost is replaced by the stretch with
respect to the metric. They obtain tight bounds on the price of
anarchy and show that the problem of deciding the existence of pure
Nash equilibria is NP-hard. Network formation under the requirement for bilateral consent for building links is studied in~\cite{corbo05}. \cite{Laoutaris2007:SNS} presents an experimental study of network formation games involving non-unit link lengths.

Network formation games have also been studied in the context of
Internet inter-domain routing.  A coalitional game-theoretic problem
modeling of BGP is introduced in
\cite{papadimitriou01algorithms} and studied further in
\cite{markakis03core}.  Also related is the work on designing strategy-proof mechanisms for
BGP~\cite{feigenbaum02bgpbased} as well as the recent work on
strategic network formation through AS-level
contracts~\cite{Anshelevich06:FOCS}.

\section{Preliminaries and Problem Definitions}

Consider a set of nodes $V=\{v_1, v_2, ..., v_n\}$. Each node
$v_i\in V$ is equipped with: (1) a link-budget $k_i$, specifying
the maximum number of nodes to which $v_i$ can establish outgoing
directed links (or wires), and (2) weights $w_{ij}$, indicating
$v_i$'s preference for communicating messages to $v_j$ (for
convenience $w_{ii}=0$). Node $v_i$ establishes a wiring
$s_i=\{v_{i_1},v_{i_2},\ldots,v_{i_{k_i}}\}$ by connecting each
one of its $k_i$ links to a different node. A global wiring
$s=\{s_1,s_2,\ldots,s_n\}$ is taken from the superposition of the
individual wirings of all the nodes, defining essentially an
edge-set of a graph with vertex set $V$. The cost for node $v_i$
under a global wiring $s$ is taken by a weighted (by preference)
summation of its distances to all destinations, i.e.,
$c_i(s)=\sum_{\forall j} w_{ij} \cdot d_s(v_i,v_j)$, where
$d_s(v_i,v_j)$ denotes the length of a shortest directed path from
$v_i$ to $v_j$ over the global wiring $s$ ($d_s(v_i,v_j)=M \gg n$
if such a directed path does not exist).

\begin{definition}\label{def:BDNFgame}
(BDNF game) The Bounded Degree Network Formation game is defined
by the tuple $\langle V, \{S_i\},\{c_i\} \rangle$, where:
\begin{itemize}
\item $V$ is a set of $n$ players, which in this case amount to the nodes of a graph.
\item $\{S_i\}$ is the set of strategies available
to the individual players. $S_i$ is the set of strategies
available to $v_i$. Strategies correspond to wirings. A game in
which each player $v_i$ can select any one of the ${n \choose
k_i}$ possible wirings is called symmetric. An asymmetric game is
one in which node $v_i$ is allowed to select nodes from a subset
$V_i\subseteq V$, and therefore may have fewer than ${n \choose
k_i}$ strategies.
\item $\{c_i\}$ is the set of cost functions for the individual players.
The cost of player $v_i$ under an outcome $s$, which in this case
is a global wiring, is $c_i(s)$.
\end{itemize}
\end{definition}
 We say that a wiring $s$ is stable if it is a pure Nash
equilibrium for the BDNF game.  In the rest of the paper we focus
on BDNF games with the following two characteristics. (1) All direct
links have unit weight\footnote{Such a distance model implies that
the end-to-end delay owes to relaying at the intermediate nodes and
not due to the crossing of the actual links. Such is the case when
the relay nodes are rather slow whereas the links themselves are
fast, which is a plausible model for P2P file-sharing and overlay
routing applications that run on software over non-specialized
hardware. In~\cite{Laoutaris2007:SNS}, the authors have studied
experimentally the case of non-uniform link weights using synthetic
and actual measurement data.} and therefore the distance
$d_s(v_i,v_j)$ becomes equal to the number of hops on a shortest
path from $v_i$ to $v_j$ over $s$. In~\cite{Laoutaris2007:SNS} it is
shown that in this case, the best-response of node $v_i$ can be
obtained from the solution of a $k_i$-median problem on an
asymmetric distance function obtained from flipping the distance
function of the residual wiring $s_{-i}=s-\{s_i\}$. (2) All nodes
have the same link budget $k_i=k,
\forall i$. We call such games $(n,k)$-uniform when all the weights
are the same (e.g., all equal to 1) and $(n,k)$-non-uniform,
otherwise.
\junk{We first define terms in the context of the peering game. As
with all games we define this game by specifying the strategic
agents, their action sets and their utilities. The strategic
agents are nodes $V = v_1, v_2, ..., v_n$ in a graph $G$. Each
node $v_i$ has a budget $b_i$ of links it can use to connect to
other nodes. The set of actions of a node $v_i$ is the collection
of all subsets of size $b_i$ of $V - {v}$. When $v_i$ selects the
action ${v_{i_1}, v_{i_2}, ..., v_{b_i}}$ this means that, in $G$,
$v_i$ is connected to ${v_{i_1}, v_{i_2}, ..., v_{b_i}}$. Links
can be considered to be directed or undirected, but in the rest of
this paper we consider only directed links. So this means that
there are arcs directed from $v_i$ to ${v_{i_1}, v_{i_2}, ...,
v_{b_i}}$. Once each node selects its action the result is the
graph $G$.
 In the rest of the paper we also assume that all the budgets
have the same value $k$, i.e. $\forall_i b_i = k$. We refer to
peering games with budget $k$ as $k$-budget games. In the uniform
peering game each node wishes to minimize its sum of distances to
all other nodes in $G$, and selects it actions accordingly. In the
nonuniform peering game each node $v_i$ has a set of weights
$w^i_j, j \neq i$ for each node $v_j$ and aims to minimize the
weighted sum of distances to all the other nodes in $G$, with the
distance to $v_j$ multiplied by the weight $w^i_j$. This
(weighted) sum of distances from a node is also referred to as the
utility of that node. We say a graph is stable if it is a pure
Nash equilibrium. We define a node to be additively $x$-stable if
it cannot improve its utility by more than $x$ additively.
Analogously we define multiplicatively $x$-stable. We define a
graph to be additively (multiplicatively) $x$-stable if no node
can improve its utility by more than $x$ additively
(multiplicatively). We define a node to be happy if its utility is
the best it can possibly be. Analogous to the definitions of
$x$-stability which compare to the existing utility we define the
notions of $x$-happy by comparing to the optimal utility. }

\section{Nonuniform Games}\label{sec:nonuniform}
In this section we ask whether pure Nash equilibria exist for all
instances of the non-uniform game. We start with asymmetric
non-uniform games, in which a node is allowed to select neighbors
from a subset of $V$, with this subset being different for
different nodes in the general case. We show that pure Nash
equilibria may not exist in this case. Then we argue that given an
asymmetric non-uniform game with no pure Nash equilibria, we can
construct an equivalent symmetric non-uniform game, thereby
establishing that symmetric non-uniform games may not have pure
Nash equilibria as well.

\begin{lemma}\label{lemma:asymmetric}
There exist instances of the asymmetric $(n,k)$-non-uniform game,
for all $n \geq 11$ and $k \geq 1$ such that they have no pure
Nash equilibria.
\end{lemma}
\begin{proof}
It is sufficient to prove the desired claim for $(11,1)$-non-uniform
games since for $n > 11$ or $k > 2$, the result follows from the
$(11,1)$-non-uniform case by just forcing all of remaining wirings to
link to specific nodes, using appropriate weights.

The basic idea is to construct an instance of an asymmetric
$(11,1)$-non-uniform BDNF game that encodes the pay-off structure of a
``matching pennies'' game~\cite{osborne94game}. To construct such an
instance we define an object that we call the ``Gadget'' (see
Fig.~\ref{fig:gadget} for an illustration). Our Gadget is made out of
two sub-gadgets, sub-gadget0 and sub-gadget1. Sub-gadget0 consists of
five nodes: a central one (0C), two bottom ones (left, 0LB and right,
0RB), and two top ones (left, 0LT and right, 0RT). Similarly for
subgadget1.

\begin{figure}[htb]
\begin{center}
\includegraphics[width=2.5in]{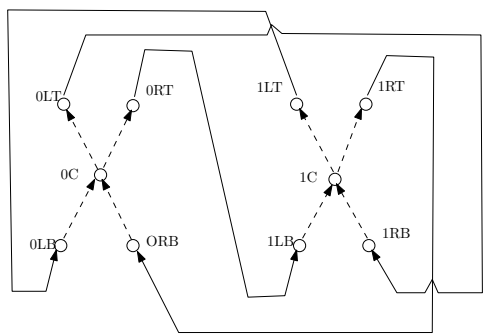}
\caption{Gadget - consisting of two subgadgets}
\end{center}
\label{fig:gadget}
\end{figure}

The solid lines in Fig.~\ref{fig:gadget} are used for originating
nodes that can connect only to a single other node. Such is the
case for the two top nodes of each sub-gadget that can connect
only to a specific bottom node of the other sub-gadget. Node 0LT
can connect only to node 1RB, node 0RT only to node 1LB, and
similarly for the other sub-gadget.

The dashed links in the same figure represent choices the
originating node can make. The central node of a sub-gadget can
connect to only one of the top nodes of the same sub-gadget. So
for example 0C can connect its single outgoing link to either 0LT
or to 0RT. Each one of the two bottom nodes of a sub-gadget can
connect to either the central node of the same sub-gadget or to
another node not depicted in the figure.

Having defined the players and the asymmetric strategy sets we
move on to define the weights $w_{vu}$. The central node in each
sub-gadget is only interested in reaching the central node of the
other sub-gadget, i.e., 0C wishes to reach 1C and vice versa,
neither one, however, is allowed to establish a direct link to the
other, and therefore they have to go first through a top node of the same sub-gadget and then through
a bottom node of the other sub-gadget.
The bottom nodes are primarily interested in reaching the
cross-over top node in the same sub-gadget, i.e., 0LB is
interested in reaching 0RT. Now, if a bottom node cannot reach its
corresponding cross over top node, then it prefers to connect to
the node that is not shown in the figure. The top nodes only care to
reach the unique node they are allowed to connect
in the other sub-gadget.

We claim that this gadget has no pure Nash equilibria. Let 0C be
connected to say 0LT and let 1C be connected to 1RT. Then it is
clear how the bottom nodes will connect: 0RB will connect to 0C,
0LB will connect to the not depicted node, 1LB will connect to 1C,
and 1RB will connect to the not depicted node. But now 1C can
reach 0C while 0C cannot reach 1C. Hence 0C is unstable and has to
switch from 0LT to 0RT, which gives it a path for reaching 1C.
This has the following immediate effects on the bottom nodes of
sub-gadget0: 0LB switches to connecting to 0C and 0RB switches to
connecting to the not depicted node. This, however, breaks 1C's
path to 0C, causing it to switch to 1LT. Now, the bottom nodes of
sub-gadget1 switch accordingly, thereby breaking 0C's path to 1C
and causing it to switch back to its initial connection to 0LT.
The bottom nodes of sub-gadget0 return to their initial positions,
thereby breaking 1C's path to 0C, and causing it to switch back to
its original connection to 1RT. This causes the bottom nodes of
sub-gadget1 to return to their initial positions. At this point
the gadget has return to its initial configuration. Thus it is
easy to see that like puppies chasing their tails the two
sub-gadgets will be continually switching. Hence there is no pure
Nash equilibrium.
\end{proof}

\begin{lemma}
There exist instances of the symmetric $(n,k)$-non-uniform game
with $n\geq 11$ and $k\geq 1$ such that they have no pure Nash
equilibria.
\end{lemma}
\begin{proof}
As in the case of Lemma~\ref{lemma:asymmetric}, it is sufficient
to prove the desired claim for $(11,1)$-non-uniform games.  We
will construct an instance of a symmetric $(11,1)$-non-uniform
game in which, although the nodes are allowed to select neighbors
from the entire node set $V$, in a pure Nash equilibrium they
would have to connect to neighbors belonging to the same subsets
as with the asymmetric $(n,k)$-non-uniform game of
Lemma~\ref{lemma:asymmetric}. We will implement this property
using the weights $w_{v,u}$. In this spirit, the solid line that
connects a node $v$ from the top of a sub-gadget to a fixed node
$u$ at the bottom of the other sub-gadget (as shown in
Fig.~\ref{fig:gadget}) can easily be implemented by setting
$w_{v,u}=\delta>0$ and zeroing out the weights to all other nodes.
Similarly, the ``switch'' from the central node $v$ of a
sub-gadget to the two top nodes $u\in\{LT,RT\}$ of the same
sub-gadget can be implemented by setting: (1) $w_{v,u}=\zeta>0$,
(2) $w_{v,v'}=\xi>0$, with $\xi<\zeta$, where $v'$ indicates the
central node of the other gadget, and (3) zeroing out the weights
to all other nodes.

Implementing the switch from a bottom node $v$ to either the
central node $u$ of the same sub-gadget, or to another node $y$
(not depicted in Fig.~\ref{fig:gadget}) is a little more involved.
Let's set $w_{v,y}=\alpha$, $w_{v,u}=\beta$, and
$w_{v,v'}=\gamma$, where $v'$ denotes $v$'s cross-over node at the
top of the same sub-gadget. If $M$ denotes the disconnection cost
we need to enforce that:
\[
\begin{array}{ll}
\alpha>\gamma \\
\alpha>\beta \\
\alpha \cdot (M-1) < \beta \cdot (M-1)+\gamma
\cdot (M-2)
\end{array}
\]
The first inequality guarantees that a bottom node will never
establish a direct link to its cross-over node at the top of the
same sub-gadget. The second one guarantees that if the link from
the central node to the cross-over does not exist, then the bottom
node will connect to the not depicted node. The last inequality
guarantees that if the link from the central node to the
cross-over node exists, the bottom node will connect to the
central node. The three inequalities can be jointly satisfied by
picking positives $\gamma,\epsilon$:
$\epsilon<\frac{M-2}{M-1}\cdot \gamma$, and setting:
$\beta=\gamma+\epsilon$ and $\alpha=\beta+\frac{M-2}{M-1}\cdot
\gamma-\epsilon$. We have now showed that there exists an instance
of symmetric non-uniform game that if it had a pure Nash
equilibrium, the asymmetric non-uniform instance of
Lemma~\ref{lemma:asymmetric} would also have one.  Since the last
lemma establishes that such a pure Nash does not exist, it follows
that there exist symmetric non-uniform games with no pure Nash
equilibria.
\end{proof}

\begin{theorem}
Given a non-uniform instance of the game
(symmetric or asymmetric) it is NP-complete to determine whether
the instance has a pure Nash equilibrium.
\end{theorem}
\begin{proof}
Proof sketch: The proof is by reduction from 3SAT with $n$
variables and $m$ clauses. The basic idea is to model each
variable as a sub-gadget (represented as a circle in
Figure~\ref{fig:npcomplete}) with the state where $0$C is pointed
to $0$LT representing the variable is set to 1 and $0$C is pointed
to $0$RT representing that the variable is set to 0. Variables are
indifferent between their two states. Each clause is represented
by a vertex that can pick one of its 3 literals to connect to. A
literal in a clause is represented by a connection, through an
intermediate vertex, that is made to the corresponding variable's
sub-gadget if and only if the sub-gadget is oriented in the right
way, i.e., to guarantee that the variable is satisfied. There is a
base node that connects to all the clauses. Finally we have a
gadget whose central nodes are interested in connecting to the
base node if and only if all the clauses connect to their
intermediate nodes, i.e., the can all be satisfied.

\begin{figure}[htb]
\begin{center}
\includegraphics[height=2.5in]{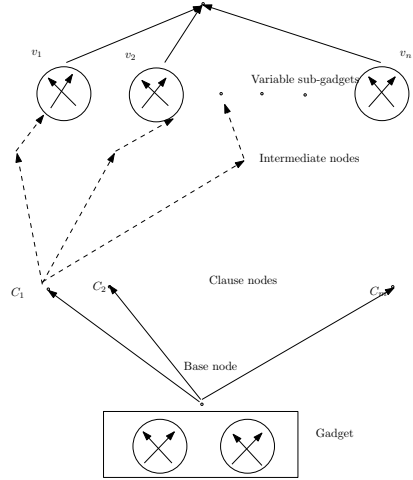}
\caption{Construction to prove NP-completeness of detecting pure
Nash Equilibrium}
\end{center}
\label{fig:npcomplete}
\end{figure}

The weights are set up so that if the formula is satisfiable
then each of the variable gadgets have an appropriate orientation
and there would be a pure Nash equilibrium. If the formula is not
satisfiable then the gadget would continually toggle.
\end{proof}

\section{Uniform Games}
\label{sec:uniform}
We counterbalance the result in Section~\ref{sec:nonuniform} by
showing that stable graphs do exist for uniform games. In
particular, we show that for every integer $n$ and any $k < n$,
the uniform $(n,k)$-BDNF game has a stable graph.  In fact, there
may be multiple stable graphs.  We first present our constructed
graph and then prove its stability. We next present in
Section~\ref{sec:uniform.structure} some structural properties of
stable graphs and argue that certain natural regular wirings, in
which all nodes imitate the same wiring pattern, are not stable.
\subsection{Description of our constructed stable graph}
Granted $k>2$ and $n>k$, let $h$ denote the maximum depth of a
full $k$-nary tree with $n_{k,h}\leq n$ nodes. Let $t=n-n_{k,h}$
denote the number of \emph{remaining} nodes, $\tau =t \mbox{ div }
k$ the number of complete $k$-tuples that can be formed based on
$t$, and $t_1=t \mod k$ the resulting \emph{untupled} nodes. We
construct the following graph:

\noindent \textbf{The tree:} We use $n_{k,h}$ nodes to construct
$T_{k,h}$, a full $k$-nary tree with $h$ levels. Nodes are
labelled according to the in-order traversal of $T_{k,h}$, with
node 1 being the root.

\noindent \textbf{The additional roots:} We consider the $t$
remaining nodes as additional roots (conceptually) placed next to
node 1 and connected to the same set of children (nodes $2 \ldots
k+1$, which we call \emph{the hubs}).

\noindent \textbf{The bridge:} We call the rightmost leaf of
$T_{k,h}$ \emph{the bridge}. We connect $t_1$ of the bridge's
links to corresponding untupled roots and the remaining $k-t_1$
ones to corresponding heaviest hubs. The weight of a hub amounts
to the number of roots connected to leafs of the subtree rooted at
the hub (minus 1 if the bridge is one of these leafs).

\noindent \textbf{The remaining leafs:} We use the $k$ links of
all the remaining leafs of $T_{k,h}$ (with the exempt of the
bridge) to connect to either an unconnected $k$-tuple of
additional roots, or the $k$ hubs. The rule for selecting which
leafs get roots and which get hubs is as follows. We start
``packing'' the $k$-tuples evenly at the leafs belonging to the
subtrees rooted at the first two children of hub node 2. If $\tau$
is odd we pack the remaining $k$-tuple to a subtree rooted at a
third child of node 2. If $\tau$ is large enough to fill all the
leafs of the first two subtrees of node 2, we continue connecting
$k$-tuples using the free links of the next available leafs of
$T[2]$. If $T[2]$ gets filled we continue likewise with $T[3]$ and
so on (minding to first balance two subtrees of $T[x]$ and then
spill to the rest of the subtree). In the next section we prove
that this construction is a pure Nash equilibrium for the BDNF
game.
\subsection{Existence of stable graphs}
\label{sec:uniform.exist}
The main result of this section is the following.
\begin{theorem}
For any $n\geq 2$ and any positive $k$, there is a uniform stable
$(n,k)$-wiring.
\end{theorem}
\begin{proof}
The theorem is clearly true for $k = 1$ because the directed
Hamiltonian cycle is stable. We first prove the theorem for $k=2$.
As it turns out, this is the hardest case. By Lemma
\ref{lemma:ravidense}, there are regular $(n,2)$-stable wirings
  for $n\leq 5$. In fact, one can easily show that there are regular $(6,2)$- and
$(7,2)$-stable wirings as well. So, we assume $n \geq  8$.

Let $h$ be the largest integer such that $2^{h} -1 \leq  n$. Let
$t = n - (2^{h}-1)$. We will consider the following cases: (0) $t=
0$, (1) $ t = 1\mod  2$, (2) $t = 0 \mod 4$ and (3) $t =2 \mod 4$.
Although $t=0$ is implied by $t = 0\mod 4$, we handle it
separately for clearity. With the assumption that $n\geq 8$, we
have $h\geq 3$.

Let us first consider the case when $t = 0$. Let $T_{2,h-1}$ be  a
complete binary tree of height $h-1$. Note that $T_{2,h-1}$ has
$2^{h}-1 = n$ node. So we label the nodes of $T_{2,h-1}$ from
$[1:n]$ according to the in-order traversal of the tree. Let $r=1$
be the root. See Figure \ref{fig:stable}. Note that the root $r$
achieves the ``utopian'' cost, which is as good  as  it can be
obtained even it controls all wiring of the links.
\begin{figure}[htb]
\begin{center}
\includegraphics[width=2.5in]{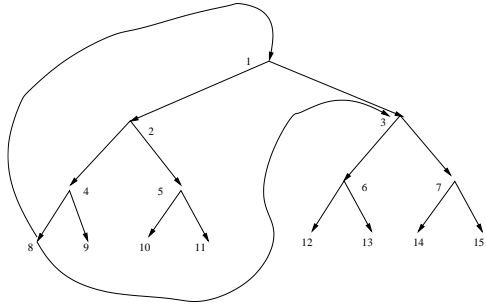}
\caption{An example of a stable wiring} \label{fig:stable}
\end{center}
\end{figure}
We choose the leftmost leaf, the node with lable $2^{h-1}$, and
use one of its out-link to connect to the root $r$. We connect its
other outlink to the right children, node 3, of the root. We
connect all other leaves directly with 2 and 3, the two children
of the root. We now argue that this wiring is stable. Clearly, the
root is stable, simply because it has a utopian cost. All internal
nodes are stable for the reason that any rewire will disconnect it
from the nodes in the subtree it no longer connects to.

Consider node $2^{h-1}$. We will call it the bridge node. Rewiring
its link to  the root will disconnect it from the root. It
suffices to show that it can not rewire its other outlink. We
first observe that a rewiring to 2 is worse because it will
increase the cost of the bridge node by 1: its distance to  every
node in $T[3]$, the subtree rooted at 3 increases by 1 and its
distance to every node in $T[2]$ other than itself decreases by 1.
Consider a rewiring of this link to a children of 3, say, its left
children 6 (recall we assume $n\geq 8$). Its distances to the
root, and to every node in $T[2]$ remains the same. Its distances
to 3 and to every node in its right subtree increase by 1, and its
distances to every node in the left subtree of 3 decrease by 1. So
the net loss is 1. Recursively, with the same argument, we can
easily show that a rewiring to any node in the subtree of 3
induces a loss. Similarly, we can show that a rewiring to any
nodes in $T[2]$ is worse than the rewiring to 2 itself. So, 3 is
the best choice of $2^{h-1}$. For other leaves, their total
distance is equal to the utopian + 1. Since no rewiring will make
them utopian, they are stable as well.

To prove the other three cases, the key is to view the wiring
correctly. We now discuss our ``correct'' view. Recall $t = n -
(2^{h}-1)$. We will view nodes $[2^{h}:n]$ and 1 all as roots.
Note that $[2^{h}:n] = [n-t+1:n]$. Both $T[2]$ and $T[3]$ are
complete binary tree of height h-2. We connect all these roots to
2 and 3. All roots are in the state of the utopian. See Figure
\ref{fig:generalStable}. In other words, whereas in the previous
Case 0, we had one root, now we have $t+1$. Let $L[2]$, $R[2]$,
$L[3]$, and $R[3]$, respectively denote the subtree of height
$h-3$, rooted at the children of 2 and 3.
\begin{figure}[htb]
\begin{center}
\includegraphics[width=3.5in]{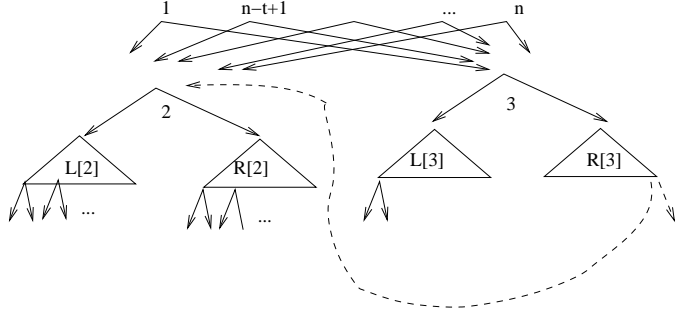}
\caption{General stable wiring} \label{fig:generalStable}
\end{center}
\end{figure}

The rule for our wiring is very simple! Case 1, when $t+1 = 0 \mod
2$, is the easiest case. We pair up these roots and suppose we
have $p$ pairs. We can choose any $p$ leaves and uses their
out-links to connect to these pairs of roots. The rest of the
leaves are connected to 2 and 3. Clearly, all roots are stable.
All leaves that are not connected to roots have cost utopian +1,
so they are stable. All other nodes are ``internal'' and they are
stable because any  local rewiring will disconnect them from some
nodes.

In Case 2, we have $t+1 = 1 \mod  4$. Suppose $t = 4p$. We form
$2p$ pairs of roots with root $n$ unpaired. If $2p \leq 2^{h-3}$,
the number of leaves in $T[2]$, we choose $p$ leaves from $L[2]$
and $p$ leaves from $R[2]$ uses their out-links to connect to
these pairs. If $2p > 2^{h-3}$,  then we choose all leaves in
$T[2]$ and also choose $(2p-2^{h-3})/2$ leaves from $L[3]$ and
$R[3]$. We use their out-links to connect to these pairs of roots.
Finally, we select a not-yet-chosen leaf $v$ from $T[3]$. Without
loss of generality, assume $v = 2^{h}-1$. We refer to it as the
bridge node. We use one of its out-link to connect to $n$ and the
other link to connect to $2$. The rest of the leaves are connected
to 2 and 3. Again, all roots and all ``internal'' nodes including
those leaves that are connected to two roots are stable. All
non-bridge leaves have cost utopian+1, so they are stable. As for
the bridge node $v$, its link to $r$ can not be rewired. If $p =
0$, we have Case 0. So $v$ is stable. Otherwise, $p>0$, and the
subtree of $2$ connects to at least $4$ more roots than the
subtree of $3$. So for $v$, connecting to $2$ is strictly better
than connecting to $3$. Because $L[2]$ and $R[2]$ connects to the
same number of roots, it is inferior to rewire the link to other
nodes in $T[2]$. So, $v$ is stable as well.

In Case 3, we have $t+1 = 3 \mod  4$. Suppose $t - 2 = 4p$. We
form $2p$ pairs of roots with $1$, $n-1$, and $n$ unpaired. When
$p>0$, we use the same approach of Case 2 to connect to these
roots. Then, we select a leaf $v_{1}$ from $L[3]$ and a leaf
$v_{2}$ from $R[3]$. Without loss of generality, assume $v_{2} =
2^{h}-1$. We refer to it as the bridge node. We connect $v_{1}$ to
$1$ and $n-1$ and $v_{2}$ to $r_{3}$. We also connect $v_{2}$ to
$2$. With the same argument as in Case 2, we can show that all
nodes are  stable.

The most tricky case is when $p = 0$, i.e., when we have three
roots. So far, we have only found one family of stable solutions
to this case. See Figure \ref{fig:general3Root}. Other than the
three bridge nodes, all other leaves connects to 2 and 3.

\begin{figure}[htb]
\begin{center}
\includegraphics[width=3.5in]{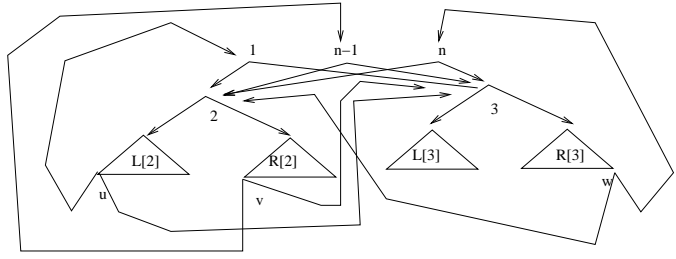}
\caption{Case with three roots} \label{fig:general3Root}
\end{center}
\end{figure}
With the same argument, we can prove that all non-bridge nodes are
stable. As bridge nodes $u$ and $v$ are symmetric, we will only
prove $u$ is stable. Clearly, it can not rewire its link to 1.
Because $T[2]$ connects to one more root than $T[3]$, and $u$
belong to $L[2]$, the wiring $(u,2)$ and $(u,3)$ induces the same
cost for $u$. If $u$ is connected to $2$, our earlier argument
will show moving the link down the tree can only increase the
cost. If $u$ is connected to $3$, because the number of roots
connected by $L[3]$ is one less than the numbers of roots
connected by $R[3]$, connecting $u$ to the root of $L[3]$ also
increases the distance to 3 by 1. So rewiring to the root of
$L[3]$ is worse than connecting to $3$. With a similar
calculation, connecting to the root of $R[3]$ induces the same
cost as connecting to $3$. Inductively, we can show $(u,3)$ is as
good as any other connection. So, $u$ is stable. Similarly, we can
show $w$ is stable.

This completes the proof for $k = 2$. For $k \geq 3$, the proof is
similar. The basic idea is start with the largest complete
$k$-nary tree that has no more than $n$ nodes. Suppose the height
of this tree is $h$. So this tree has $n_{k,h} = \sum_{i=0}^{h}
k^{i}$ nodes. Let $t = n - n_{k,h}$. We also call other $t$ nodes
roots and connect them to the $k$ children of 1, i.e., $[2:k+1]$.
Let $t_{1} = t \mod k$. We group $t-t_{1}$ roots into $k$-tuples.

We split these $k$-tuples evenly to the subtrees rooted at the
first two children of 2. These two subtrees played the role of
$L[2]$ and $R[2]$ in our proof for $k=2$. If the number of the
tuples is odd, we connect to the root in one of the tuple from a
leaf not in these two subtrees. If these two trees are full, we
uses the leaves from left to right to connect to roots.

We use the rightmost leaf  of the subtree rooted at $k+1$ as our
bridge node. The bridge node connects to $t_{1}$ roots. Other
links of the bridge nodes connects to nodes in $[2:k+1]$ in the
order of the number of leaves that their subtrees connect to. All
other leaves connects to $[2:k+1]$

Similar to the case when $k=2$, we can prove all nodes are stable.
In fact, the argument is easier, because moving a link down the
tree loses more than when $k>2$.
\end{proof}


\subsection{Properties of stable graphs}
\label{sec:uniform.structure}
We first show that all stable graphs are almost fair.
\begin{lemma}
\label{lem:fair}
In any stable graph for the $(n,k)$-uniform game, the cost of any node
is at most $n + n \lfloor \log_k n \rfloor$ more than, and at most $2
+ 1/k + o(1)$ times, the cost of any other node.
\end{lemma}
\begin{proof}
Let $G$ be a stable graph for the $(n,k)$-uniform game and let $r$ be
a node in $G$ that has the smallest cost $C^*$.  Consider the shortest
path tree $T$ rooted at $r$.  Let $v$ be any other node.  Within
$\lfloor \log_k n \rfloor$ hops from $v$, there exists a node $u$ that
has at least one edge not in $T$.  Since $G$ is stable, node $u$ has
cost at most $C^* + n$, since it can achieve this cost by attaching
one of its edges not in $T$ to $r$.  Therefore, the cost of $v$ is at
most $C^* + n + n\log_k n$, since the distance from $v$ to any node
$w$ is at most $\log_k n$ more than that of $u$ to $w$.  Noting that
$C^*$ is at least $\sum_{0 \le i < \log_k n} ik^i \ge (n - n/k)\lfloor
\log_k n \rfloor$ completes the proof of the lemma.
\end{proof}

We next present an upper bound on the diameter of any stable graph
for an $(n,k)$-uniform game.  For $k = 1$, it is trivial to see
that when $k=1$ then the simple directed cycle is the unique
stable graph (up to isomorphism), yielding a diameter of $n$.

\begin{lemma}
\label{lem:diameter}
For any $k \ge 2$, the diameter of any stable graph for an $n$-node
$k$-uniform graph is $O(\sqrt{n \log_k n})$.  Furthermore, in any
stable graph, the distance from a node with minimum cost to any node
is $O(\sqrt{n})$.
\end{lemma}
\begin{proof}
Let $G$ be a stable graph for the $(n,k)$-uniform game, and let
$\Delta$ denote the diameter of $G$, given by a path from a node
$r$ to a node $v$. Consider a shortest path tree from $r$; so the
depth of this tree is $\Delta$ and $v$ is a leaf of $T$.  Let $P$
denote the set of nodes on the path from $r$ to $v$ in $T$, not
counting $r$; so $|P| = \Delta$.  Let $C$ be the sum of distances
from $r$ to the $n-\Delta$ nodes not in $P$.  The sum of distances
from $r$ to the $\Delta$ nodes in $P$ is exactly
$\Delta(\Delta+1)/2$. So the cost of $r$ is $C +
\Delta(\Delta+1)/2$.

The cost of $v$ is at most $C + n - \Delta/2 + \Delta(\Delta/2 + 1)/4
+ \Delta(\Delta/2 + 1)/4$ since $v$ can use one of its at least two
edges to connect to $r$ and the other to connect to a node halfway
along the path from $r$ to $v$.  Simplifying, we obtain that the cost
of $v$ is at most $C + n + \Delta^2/4$.  By Lemma~\ref{lem:fair}, the
cost of $v$ is at least $C + \Delta(\Delta+1)/2 - n - n \log_k n$.  We
thus obtain the inequality:
\[
C + n + \Delta^2/4 \ge C + \Delta(\Delta+1)/2 - n - n \log_k n,
\]
yielding that $\Delta = O(\sqrt{n\log_k n + 2n})$.

The second part of the lemma can be proved using the same argument as
above with the modification that instead of invoking
Lemma~\ref{lem:fair}, we have a lower bound of $C$ for the cost of
$v$.
\junk{
Now, since $D > \log_{k-2}n$ there must be a node, say
$s$, within distance $\log_{k-2}n$ of $r$ that has (out)degree at most
$k-2$ in the tree. Since the configuration is a Nash Equilibrium the
cost of $s$ is at most the cost obtained by connecting its two "free"
links, one to $r$ and another to a point halfway on the diameter. In
this case the cost for $n$ is $n-D + C +
\frac{D^2}{4}$. And so the cost of the tree rooted at $r$ consisting
of going to $s$ and then picking up the shortest path tree rooted at
$s$ must be higher than that of the current shortest path tree rooted
at $r$, i.e. $n\log_{k-2}n + n - D + C + \frac{D^2}{4} > C +
\frac{D^2}{2}$ i.e. $D = O(\sqrt(n\log_{k-2}n)$.
}
\end{proof}

A natural degree-$k$ wiring to consider is to map the nodes to
$Z_n = \{0,1,\ldots,n-1\}$ and have the $k$ edges for all nodes be
defined by $k$ offsets $a_0$, $0 \le i < k$: the $i$th edge from
node $x$ goes to $x + a_i \bmod n$.  We refer to such wirings as
{\em regular wirings}. For a suitable choice of the offsets, these
graphs have diameter $O(n^{1/k})$.  In this section, we study a
more general class of wirings that includes regular wirings ---
namely Abelian Cayley graphs --- and show that these graphs are
not stable for $k \ge 2$.  Cayley graphs have been widely studied
by mathematicians and computer scientists, and arise in several
applications including expanders and interconnection networks
(e.g.,
see~\cite{alon+roichman:cayley,annexstein+br:group,cooperman+fs:cayley}).

A Cayley graph $G(H,S)$ is defined by a group $H$ and a subset $S$ of
$k$ elements of $H$.  The elements of $H$ form the nodes in $G$, and
we have an edge $(u,v)$ in $G$ if and only if there exists an element
$a$ in $S$ such that $u \cdot a = v$, where $\cdot$ is the group
operation.  A Cayley graph $G(H,S)$ is referred to as an Abelian
Cayley graph if $H$ is Abelian (that is, the operation $\cdot$ is
commutative).  The regular wiring described in the preceding graph is
exactly the Cayley graph with the group $H$ being the Abelian additive
group $Z_n$ and $S = \{a_i \bmod n: 0 \le i \le k\}$.

Our proof of the non-existence of pure Nash equilibria in Abelian
Cayley graphs is based on a particular embedding of these graphs into
$k$-dimensional grids.  Let $G(H,S)$ be a given Abelian Cayley graph
and let the $k$ elements of $S$ be $a_i$, $0 \le i < k$.  We assume
without loss of generality that $S$ does not contain the identity of
$H$ since these edges only form self-loops, which clearly cannot
belong to any stable graph.  Each edge of the graph $G$ can be labeled
by the index of the element of $S$ that creates it; that is, if $v = u
\cdot a_i$, then we call the edge $(u,v)$ an {\em $i$-edge}.  The edge
labels naturally induce labels on paths as follows.  If a path
contains $x_i$ $i$-edges, then we label the path by the vector
$\vec{x} = (x_1, \ldots, x_i, \ldots, x_k)$.  Note that the length of
a path with label $\vec{x}$ is simply $\sum_{1
\le i \le k} x_i$.  Furthermore, the commutativity of the underlying group
operator implies that for all nodes $v$ and all path labels $\vec{x}$,
every path that starts from $v$ and has label $\vec{x}$ ends at the
same node.

We say that node $v$ has label $\vec{x}$ if there exists a shortest
path from $r$ to $v$ that has label $\vec{x}$.  For any node $v$,
while two shortest paths from $r$ to $v$ share the same sum of
label-coordinates, the actual path labels may be different; therefore,
a node may have multiple labels.  However, a particular label is
assigned to at most one node.

We are now ready to prove that for $k \ge 2$ no Abelian Cayley
graph is stable.  For $k = 1$, it is trivial to see that the
simple directed cycle is an Abelian Cayley graph and is stable.

\begin{theorem}
For any $k \ge 2$, no Abelian Cayley graph with degree $k$ and $n$
nodes is stable, for $n \ge c 2^k$, for a suitably large constant $c$.
\end{theorem}
\begin{proof}
We now consider the impact of replacing the $i$-edge from root $r$ to
$r_i = r \cdot a_i$ by the edge from $r$ to $r'_i = r \cdot a_i \cdot
a_i$.  The node $r$ equals $(0,0,\ldots,0)$, while the node $r_i$
equals $(0,0,\ldots,1,\ldots,0)$ with a $1$ in the $i$th coordinate.
(We note that $r$ and $r_i$ are distinct since $a_i$ is not identity.)
For every node $v$ that has a label $\vec{v}$ such that $v_i \ge 2$,
the distance decreases by $1$.  Let $S_i = \{v: v \mbox{ has a label
 }\vec{v} \mbox{ with }v_i \ge 2\}$ be the set of such nodes.  On the other
hand, the only node whose distance from $r$ increases is the node
$r_i$; this is because any path in the original graph starting from
$r$, having exactly one $i$-edge $(r,r_i)$ and having length at least
two, can be substituted by another path of the same length with an
$i$-edge as its second edge.

We bound the increase in the distance from $r$ to $r_i$ in terms of
the diameter $\Delta$ of the graph.  Let $w = r \cdot a_j^{-1} \neq
r_i $ denote a node that has an edge to $r$ in $G$.  Since the
shortest path to any vertex other than $r_i$ has not increased, we
obtain that the distance from $r$ to $r_i$ is at most $\Delta+2$,
given by a shortest path from $r$ to $w$, followed by an $i$-edge and
then by a $j$-edge.

We thus obtain that when the edge $(r,r_i)$ is replaced by the edge
$(r,r'_i)$, the total utility for node $r$ decreases by at least
$|S_i| - (\Delta + 2)$.  By the definition of $S_i$, we obtain that is
precisely the set
\[
G \setminus \bigcup_{0 \le i < k} S_i = \{\vec{v}: 0 \le v_i \le 1 \mbox{ for all }i
\}.
\]
Thus there exists $i$, $0 \le i < k$, such that $|S_i| \ge (n -
2^k)/k$.  Therefore, the graph $G$ is not stable if $(n - 2^k)/k$
exceeds $\Delta + 1$.

By Lemma~\ref{lem:diameter}, for $G$ to be stable $\Delta =
O(\sqrt{n})$.  We now use this upper bound on $\Delta$ to obtain that
if $n \ge c 2^k$, for an appropriately large constant $c$, then $(n -
2^k)/k$ exceeds $\Delta + 1$, implying that $G$ is not stable.
\end{proof}

\begin{corollary}
For any $k>4$, the $2^k$-node hypercube is not stable for the
$(2^k,k)$-uniform game.
\end{corollary}

If the degree $k$ is more than nearly half the size of the graph, then
one can show that any degree-$k$ $n$-node Abelian Cayley graph is
stable.
\begin{lemma}\label{lemma:ravidense}
For all $k > \frac{n-2}{2}$ any degree-$k$ $n$-node Abelian Cayley
graph is stable.
\end{lemma}

\junk{
Any node that is reachable from $\pi(r_i)$ in the shortest path tree
has its distance from $r$ decreased by 1.  By
Lemma~\ref{lem:monotone}, there are at least $\prod (x_i + 1)$ nodes
in $G$, one corresponding to each vector $y
\le x$.  Of these, at least $(x_i - 1) \prod_{j \neq i} (x_j + 1)$ are
reachable from $\pi_i(r_i)$.  On the other hand, the only node whose
distance from $r$ increases is the node $r_i$, and the increase in
distance is by $k + 1 + \sum_i x_i$.

If $x_i > 2k+3$ and there exists $x_\ell$, $\ell \neq i$, such that
$x_\ell \ge 1$, then $(x_i - 1)(x_\ell + 1) > x_i + x_\ell + 2k + 1$.
This implies that $(x_i - 1) \prod_{j \neq i} (x_j + 1)$ exceeds $k +
1 + \sum_i x_i$, yielding the desired result.

If all the $x_j$'s are less than $2k+3$, then $k + 1 + \sum_i x_i =
O(k^2)$.  On the other hand, at least one of the $k$ lines adjacent to
$(0,0,0,\ldots,0)$ has at least $n^{1/k}$ nodes.  So if $n^{1/k} - 2$
exceeds $O(k^2)$, which holds for $n$ sufficiently large, we are done.

The final case we need to consider is when $\vec{x}$ is of the form
$(0,\ldots, 0,x_i,0\ldots,0)$, where $x_i > 2k + 3$.
}

\section{Analysis of best response walks}
\label{sec:best}
Given the existence of pure Nash equilibria for $(n,k)$-uniform games,
a natural question that follows is whether an equilbrium can be
obtained by a sequence of local rewiring operations.  In particular,
we consider best response walks, in each step of which a node tests
for its stability and, if not stable, rewires its links to the set of
nodes that optimize its cost.  We assume for convenience that in any
step of the best response walk only one node attempts to rewire its
links.

We first show in Section~\ref{sec:best.strong} that starting from any
initial wiring, best response walks quickly converge to a strongly
connected network.  We next study convergence to stability in
Section~\ref{sec:best.stable}, and show that there exists an initial
state from which a particular best response walk does not converge to
a stable wiring.  We also present results of some experiments that
study convergence to stability of best response walks from regular and
random wirings.

\subsection{Strong connectivity}
\label{sec:best.strong}
In this section, we show that starting from any initial state, the
best response walk converges to a strongly connected graph in $O(n^2)$
steps, as long as every node is allowed to execute a best response
step once every $n$ steps.  Furthermore, there exists an initial state
such that a best response walk takes $\Omega(n^2)$ steps to converge
to strong connectivity.

For a given node $u$, we define the {\em reach}\/ of $u$ to be the
number of nodes to which it has paths.  Since the cost of
disconnection is assumed to be $M  \gg n$, when we execute
best-response for a node $u$, the reach of $u$ cannot decrease.
\begin{lemma}
\label{lem:reach}
Suppose the graph $G$ is not strongly connected, and a node $u$
changes its edges according to a best response step.  Then, after the
step, the reach of any node other than $u$ either remains the same or
is at least the new reach of $u$.
\end{lemma}
\begin{proof}
If a node $v$ has a path to $u$, then the reach of $v$ is at least the
reach of $u$ after the best response step.  Otherwise, the reach of
$v$ does not change.  \junk{Let $r = \reach(u)$.  Let $S_<$, $S_=$, and
$S_>$ denote the set of nodes with reach smaller than, equal to, and
greater than $r$ before the best response step.  Similarly, let
$S'_<$, $S'_=$, and $S'_>$ denote the set of nodes with reach smaller
than, equal to, and greater than $r$ after the best response step.
Then, we obtain $S'_< = S_<$ and $|S'_=| < |S_=|$ since the reach of
$u$ increases after the step and the reach of any node in $S_= \cup
S_>$ is greater than $r$ after the best response step.  Therefore,
$\Reach'$ is lexicographically larger than $\Reach$.}
\end{proof}
The above lemma indicates that whenever a best response step causes a
change, the vector that consists of all the reach values in increasing
order becomes lexicographically larger.  In order to show convergence,
we need to argue progress.  We will do so by showing that whenever the
graph is not strongly connected, then there exists a node that can
improve its reach.  We, in fact, argue a stronger property that allows
us to bound the convergence time.

We consider best response walks that operate in a round-robin manner.
Each round consists of $n$ steps; in each round, each node executes a
best response step in an arbitrary order.  (The order may vary from
round to round.)
\begin{lemma}
\label{lem:progress}
If $G_r$ is not strongly connected at the start of a round $r$, then
the minimum reach increases by at least one at the end of the round.
\end{lemma}
\begin{proof}
Consider the strongly connected components of the given graph
$G_r$. Consider the ``component graph'' $CG$ in which we have a
vertex for each strongly connected component and edge between two
components whenever there is an edge from a vertex in one
component to the other.  This graph is a dag.  Let $m$ denote the
minimum reach in $G_r$.  By Lemma~\ref{lem:reach}, nodes with
reach greater than $m$ will always have reach greater than $m$. So
we only need to consider nodes with reach $m$.  All of these nodes
lie in sink components.

Consider any sink component $C$.  We first argue that there exists
a node in $C$ that can improve its reach by executing a best
response step.  Consider a vertex $u$ in $C$ that has an edge from
a vertex $v$ in another component.  Let $w$ be a vertex in the
sink component that has an edge to $u$.  All of $u$, $v$, and $w$
exist by definition of strongly connected components (and our
assumption that the out-degree of every vertex is at least 1).  If
$w$ removes the edge $(w,u)$ and replaces it by $(w,v)$, it can
reach all vertices in the sink component as well as the component
containing $v$.  The latter set is clear; for the former set, note
that all we have done is replace the direct edge $(w,u)$ by the
two-hop path $w \rightarrow v \rightarrow u$.

For any sink component $C$, let $v$ be the first node in $C$ in
the round order that improves its reach through a best response
step. Note that $v$ exists, by the argument of the preceding
paragraph. Furthermore, in the step prior to $v$'s best response,
the reach of every node in $C$ is $m$.  After $v$'s best response,
the reach of $v$ increases to at least $m + 1$, and so does that
of every node in $C$ since they each have a path to $v$.  By
Lemma~\ref{lem:reach}, after every subsequent step, the reach of
any node in $C$ is at least $m + 1$.  Therefore, it follows that
at the end of the round, the reach of every node in a sink
component of $CG$ increases; hence, the minimum reach increases,
completing the proof of the lemma.
\end{proof}

\begin{theorem}
\label{thm:converge}
The best response walk converges to a strongly connected graph in
$n^2$ steps.
\end{theorem}
\begin{proof}
By Lemma~\ref{lem:progress}, the minimum reach increases by at least
one.  Since the initial reach is $1$ and the maximum reach is $n$, the
number of steps for the best response walk to converge to a strongly
connected graph is at most $n^2$.
\end{proof}

We next show that the above theorem is essentially tight by presenting
a scenario in which a best response walk may take $\Omega(n^2)$ steps
to converge to a strongly connected graph.  Consider a graph $G$ of $n
= r + p$ nodes that is a directed ring over $r \ge n/2$ nodes together
with a directed path of $p = n - r$ nodes that ends at one of the
nodes in the ring.  Suppose a round begins at the tail $T$ of the
directed path, which can reach all nodes, proceeds along the path and
then along the ring in the direction of the ring.  The $p$ nodes on
the path cannot improve their reach.  Furthermore, the first $r - p$
nodes on the ring (in round-robin order) also cannot improve their
reach in a best response step.  The $(r - p + 1)$st node can improve
its reach by connecting to $T$, yielding a new graph $G'$ that is a
directed ring over $r + 1$ nodes and a directed path of $n - r$ nodes.
If we repeat this process, the number of steps to converge is
$\Omega(n^2)$.
\subsection{Stability}\label{sec:best.stable}
\begin{figure}[htb]
\centering
\includegraphics[width=1.7in]{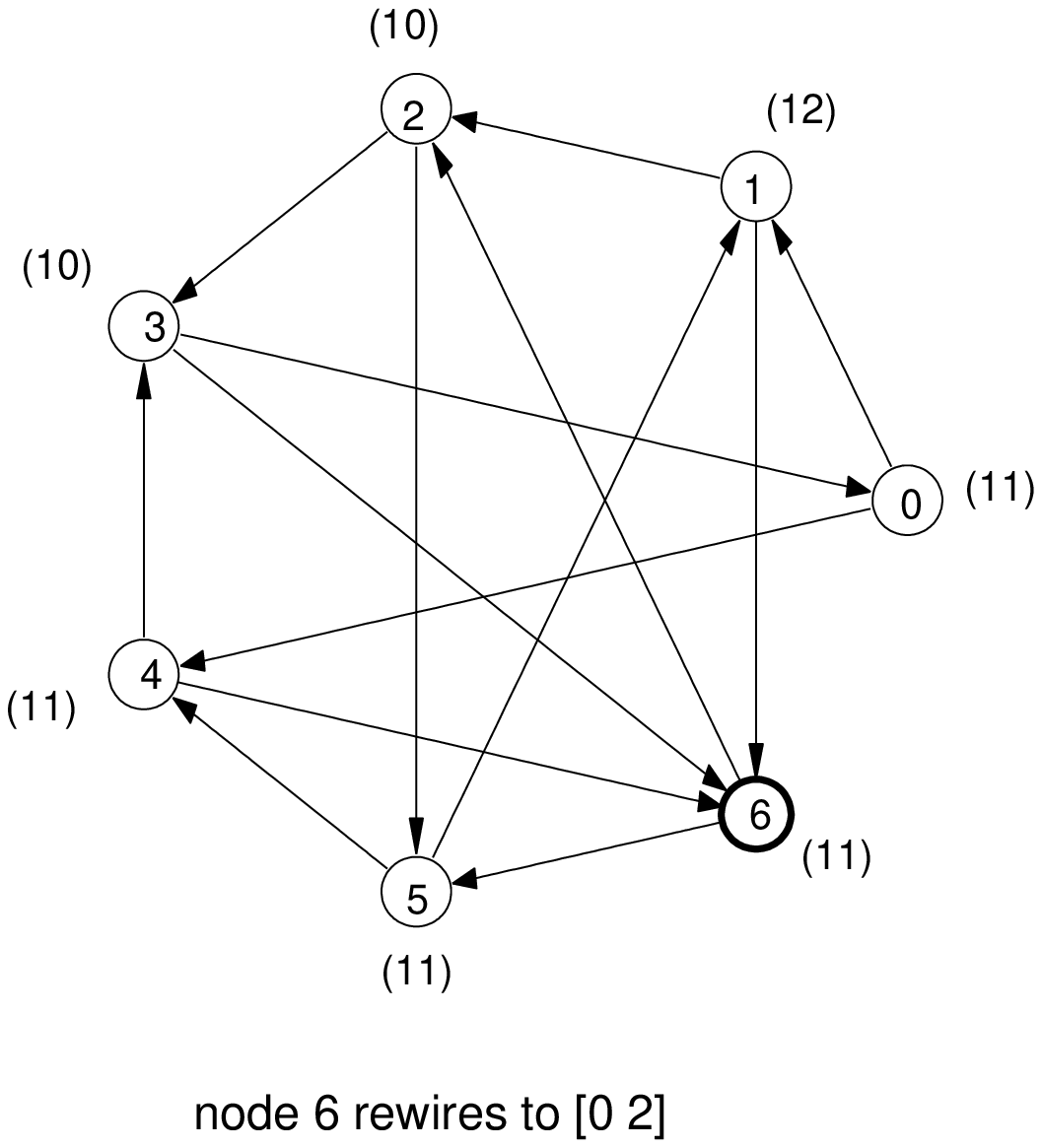}\includegraphics[width=1.7in]{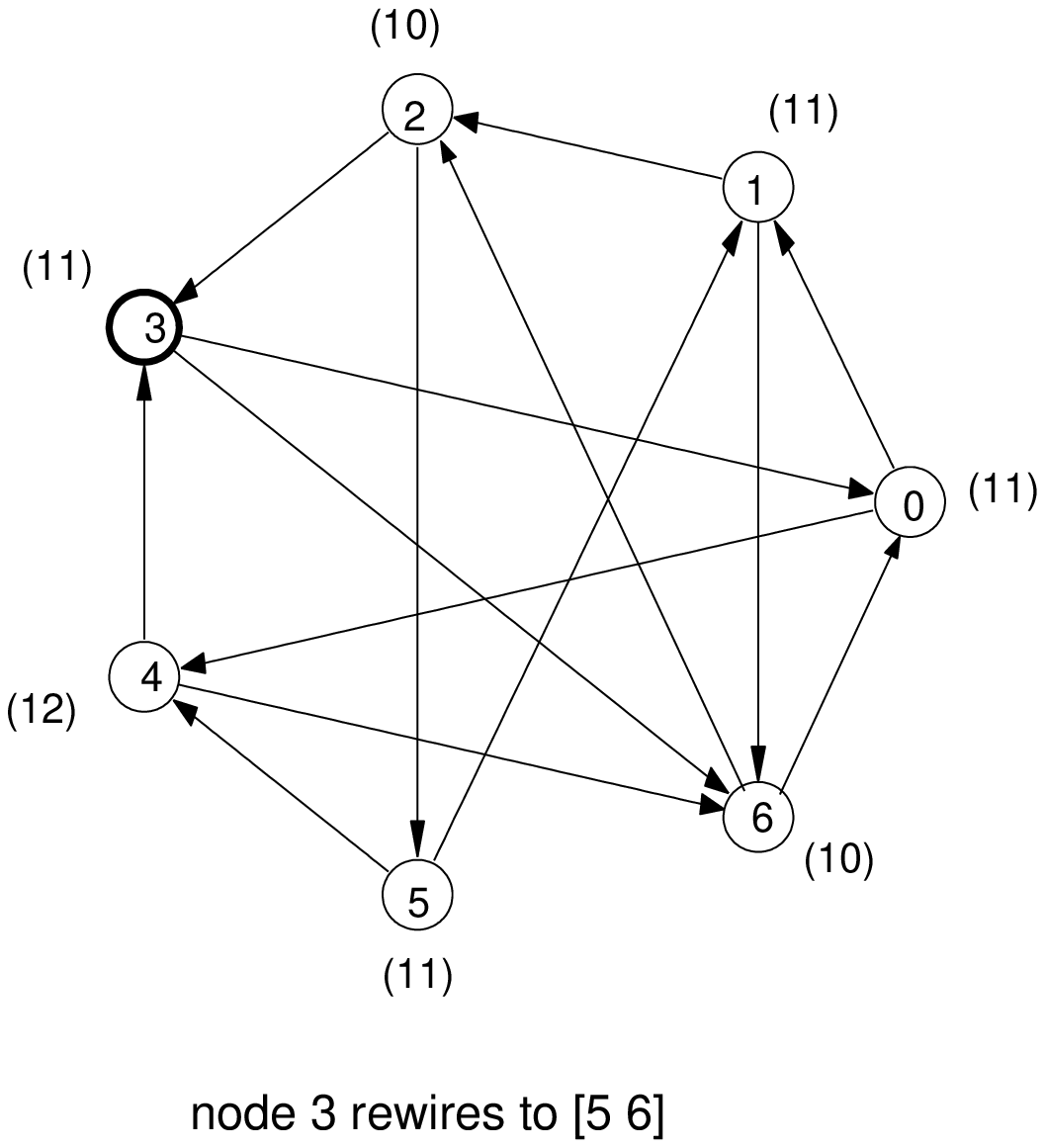}\includegraphics[width=1.7in]{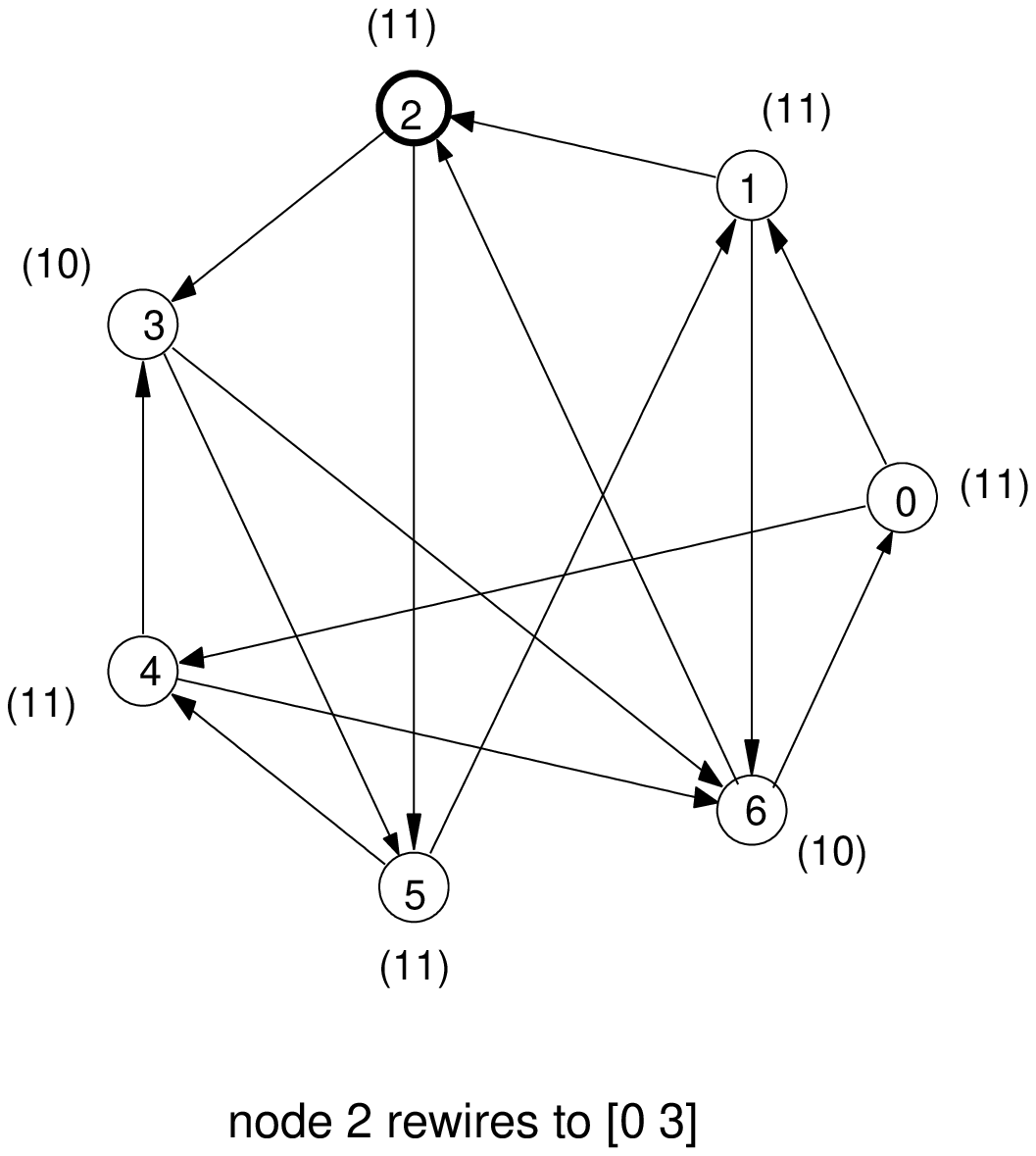}
\includegraphics[width=1.7in]{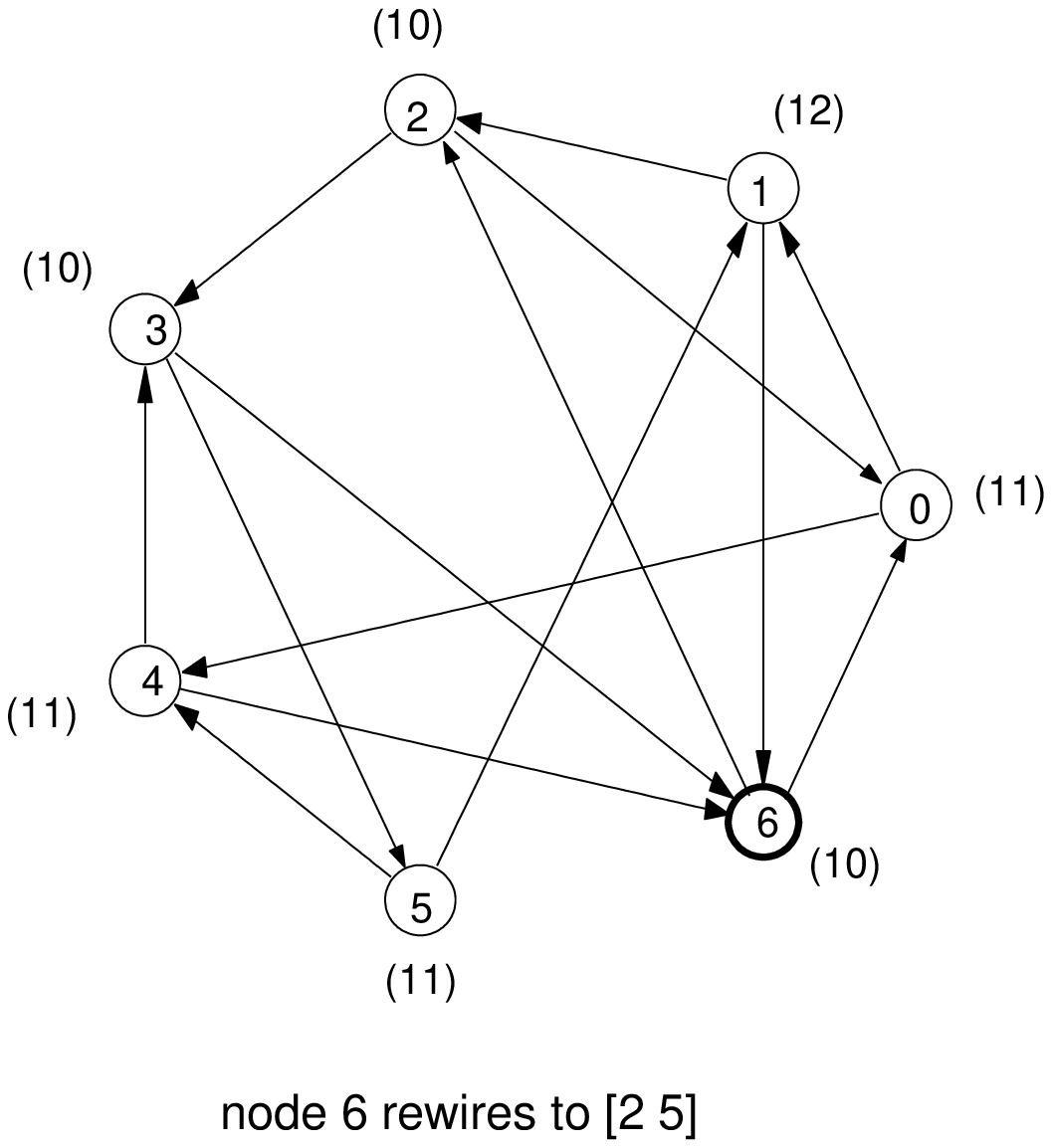} \includegraphics[width=1.7in]{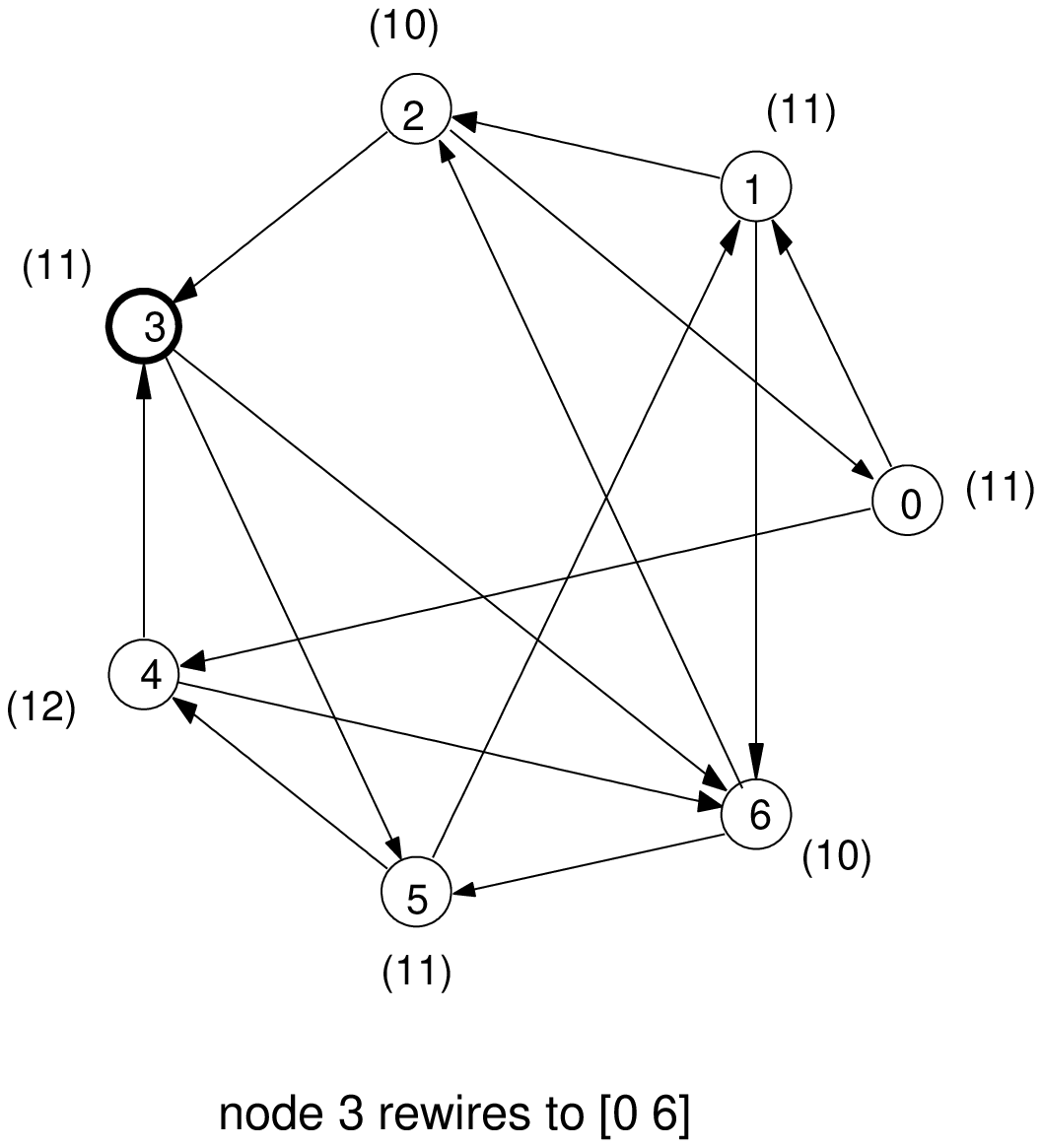}\includegraphics[width=1.7in]{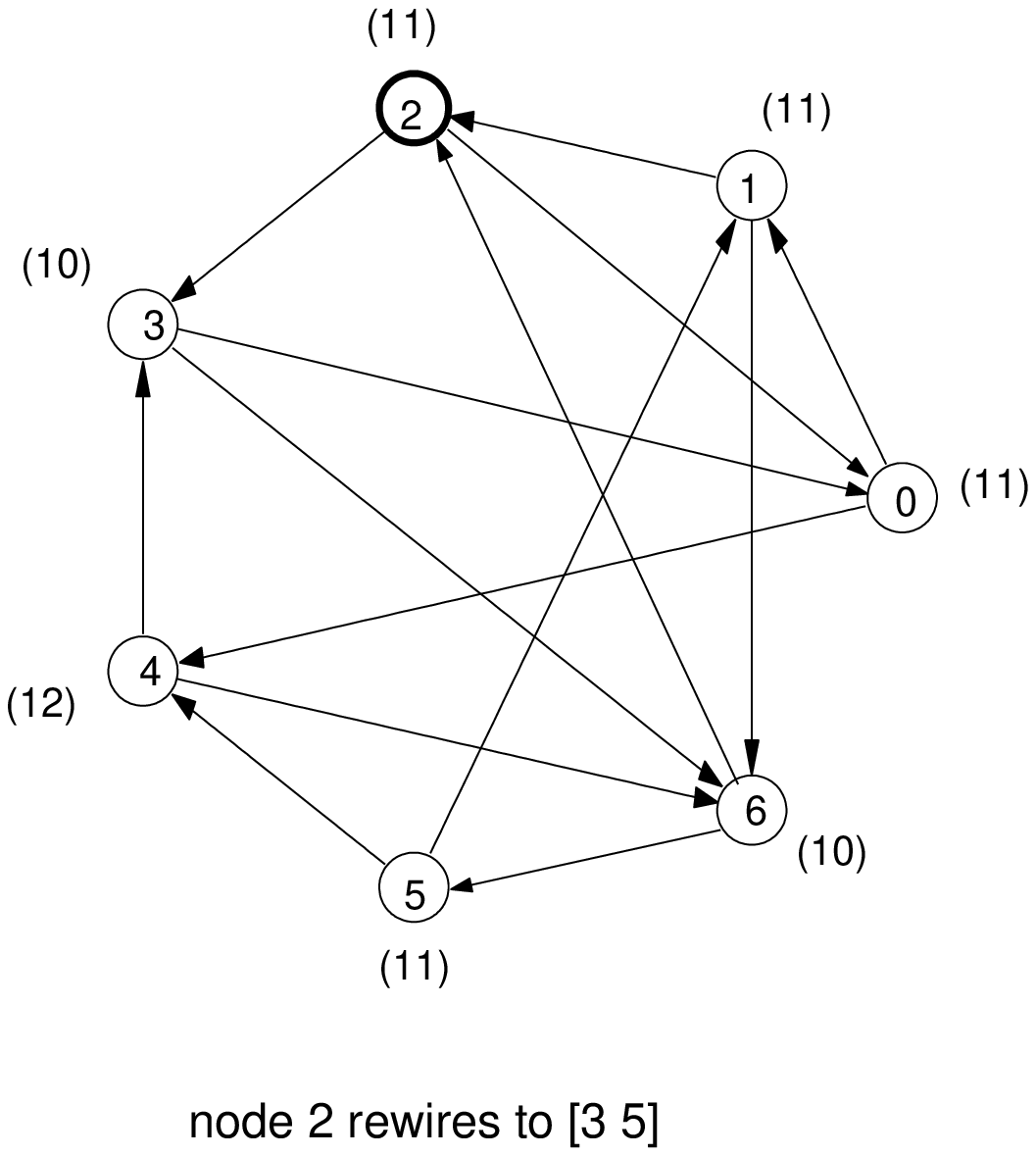}
\caption{An example in which a round-robin best-response walk
loops. Starting from the top left configuration and following a
round-robin best-response walk $6\rightarrow 0 \rightarrow 1
\rightarrow \ldots \rightarrow 6 \rightarrow 1 \ldots$ we get back
to the initial configuration after 6 deviations (nodes
$6,3,2,6,3,2$). Turns that are not illustrated imply stable nodes.
Next to each node we indicate its cost under the current
configuration.}\label{fig:looping}
\end{figure}
\begin{figure}[tb]
\centering
\includegraphics[angle=-90,width=2.8in]{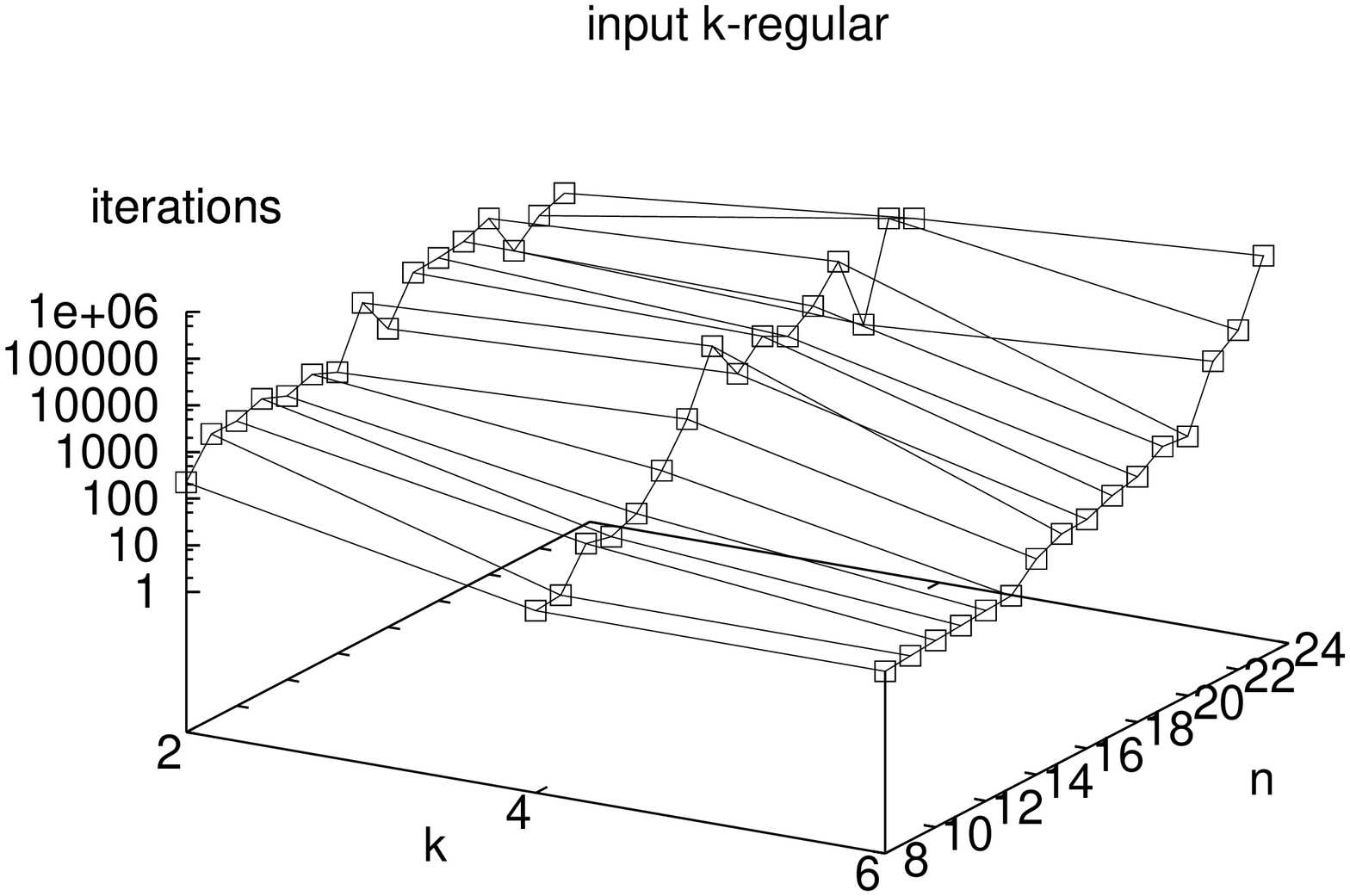}
\includegraphics[angle=-90,width=2.8in]{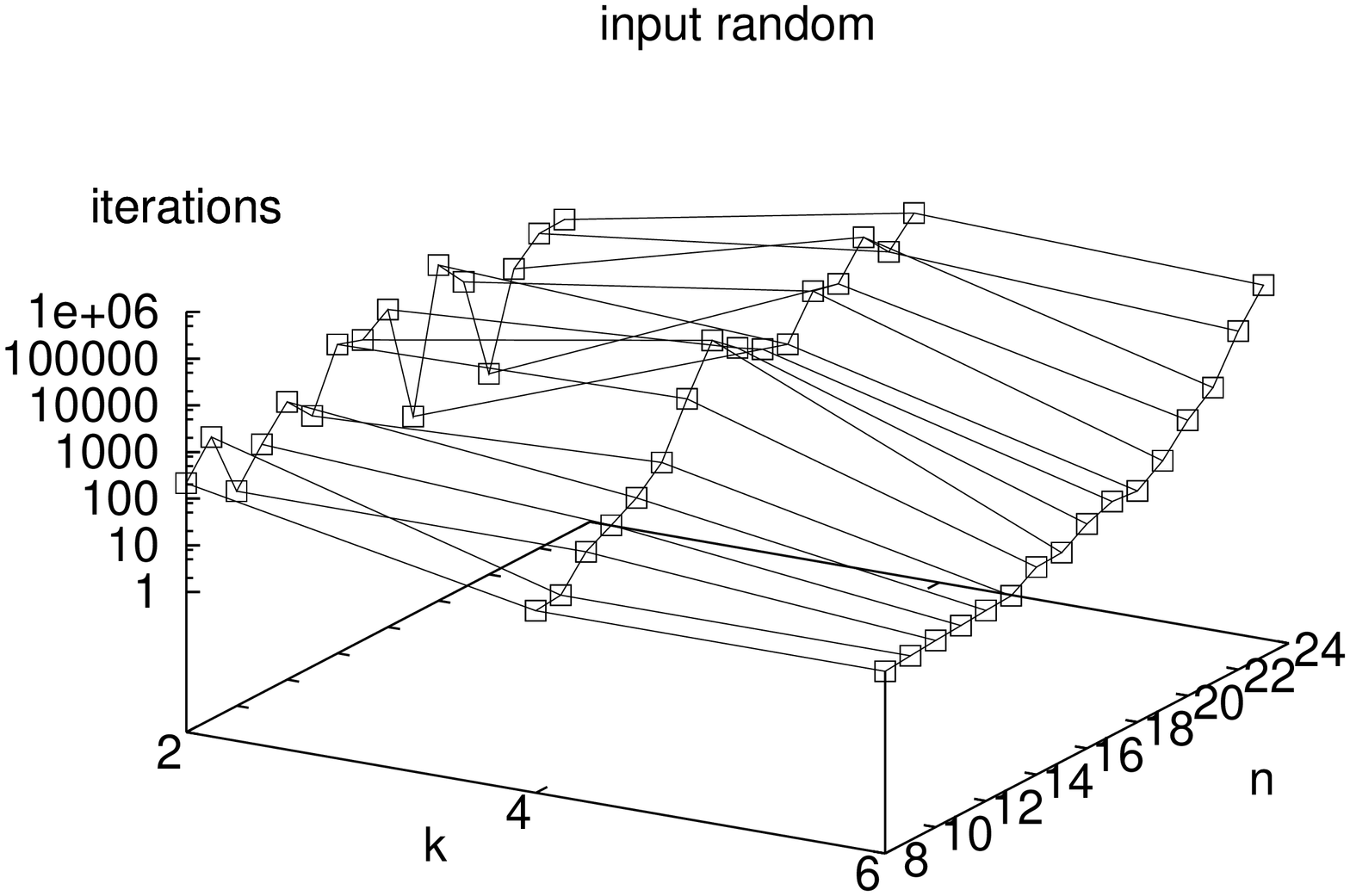}
\caption{Convergence time starting from regular and random initial
graphs.} \label{fig:experimentConv}
\end{figure}
Unlike strong connectivity, convergence to a pure Nash equilibrium is not guaranteed. Next we
show a simple example in which a round-robin best-reponse walk loops. Our simple example
is on a (7,2)-uniform game that starts from the top-left configuration of Fig.~\ref{fig:looping}. The nodes
take turns in round-robin order, starting with node 6 then nodes 0,1,2, and so on. Tracing the example, one
can verify that after 6 deviations (nodes $6,3,2,6,3,2$ re-wiring in this order, implying that missing nodes are stable), the graph
returns to the initial configuration thus completing a loop.

The above example of a loop in the best response walk, does not rule
out the possibility that either (a) a well-chosen best response walk
converges from any initial state, or (b) certain best response walks
do converge to stability if started from simple initial wirings such
as the empty graph.

In order to address the preceding questions, we have conducted
experiments on best response walks in large networks.  We have
discovered initial wirings for which the best response walk in which a
node with the maximum cost always executes the best response step does
not converge to a stable wiring.  On the other hand, in all of our
experiments, this best response walk always converges to a stable
wiring, if started from an empty wiring.

Another closely related question is: From an empty wiring (or any
initial wiring), is there a best-response walk of length polynomial in
$n$ that leads to a pure Nash equilibrium?  Our experiments suggest
that from some initial wirings there might be an exponentially long
best-response path to a pure Nash equilibrium.

In the first plot of Figure~\ref{fig:experimentConv}, we start
round-robin best-response walk from a regular $(n,k)$-wiring with
offsets $[1:k]$.  All our experiments converge to a stable wiring.
We plot the lengths of walks for all $(n,k)$-pairs.  In the second
plot of Figure~\ref{fig:experimentConv}, we repeat the same
experiment starting from a wiring constructed as follows: Starting
from a simple directed Hamiltonian cycle, we add to every vertex
$k-1$ random out-going links.  Both experiments demonstrate
lengthy and possible exponential convergence.  Moreover, the
``random'' wiring experiment shows large variance in the length of
convergence, especially for sparse wirings.

\section{Open Problems}
\label{sec:open}
In this paper, we have proved that for any $k$, the uniform Bounded
Degree Network Formation (BDNF) Game always has a pure Nash
equilibrium.  In fact, the total social cost of our equilibrium is
within a factor of 2 of the best possible network that could be
constructed by a central network designer.  In contrast, for
non-uniform $(n,k)$-BDNF games, a pure Nash equilibrium may not always
exist.  In fact, we show that deciding the existence of a pure Nash
equilibrium is an NP-hard problem.

Several interesting and fundamental questions about the Bounded Degree
Network Formation game remain open.  Resolving them will be part of
our future research.  Here, we would like to discuss some of these
questions and also state some conjectures inspired by our experiments.

\subsection*{Best-Response Paths to Pure Nash Equilibria}
The first set of questions concerns the convergence of best-response
walks.  We would like to know whether the following statement is true.

\begin{itemize}
\item [Q1] From every initial wiring  of an $(n,k)$-BDNF network,
does there exist a best-response walk that leads to stable wiring?
\end{itemize}
In Section~\ref{sec:best}, we have given an example of an initial
wiring from which some best-response walks induce a loop.  But our
example does not rule out a positive answer to Q1.  There are several
weaker statements about convergences that are worthy of further study.

\begin{itemize}
\item [Q2]
From every initial wiring of an $(n,k)$-BDNF network, does there exist
  a strictly improving response-path that leads to a stable wiring?
\item [Q3]
From an empty wiring, is there a round-robin best-response path that
  always leads to a stable wiring?
\end{itemize}
If the answer to Q1 is yes, an interesting follow-up question is the
  following.
\begin{itemize}
\item [Q4]
From every initial wiring of an $(n,k)$-BDNF network, does there exist
a best-response walk that leads to the stable graph constructed in
Section~\ref{sec:uniform.exist}?
\end{itemize}
Our experiments seem to suggest that there exists a best response walk
that takes the empty wiring to the stable graph of
Section~\ref{sec:uniform.exist} for $n \gg k$.  Note that if the
answers to both Q1 and Q4 are yes, then the problem of finding a pure
Nash equilibrium that is reachable by a best-response walk is not
PLS-hard.  Also, if best response walks indeed converge to stability,
then resolving their convergence time is an important open problem.

\junk{
Another closely related question is: From an empty wiring (or
any initial wiring), is there a best-response path of length
  polynomial in $n$ that leads to a pure Nash equilibrium.
Our experiments, see below, suggests that from some initial wirings,
  there might be an exponentially long best-response path to a pure
  Nash equilibrium.

\begin{figure}[htb]
\begin{center}
\includegraphics[angle=-90,width=3in]{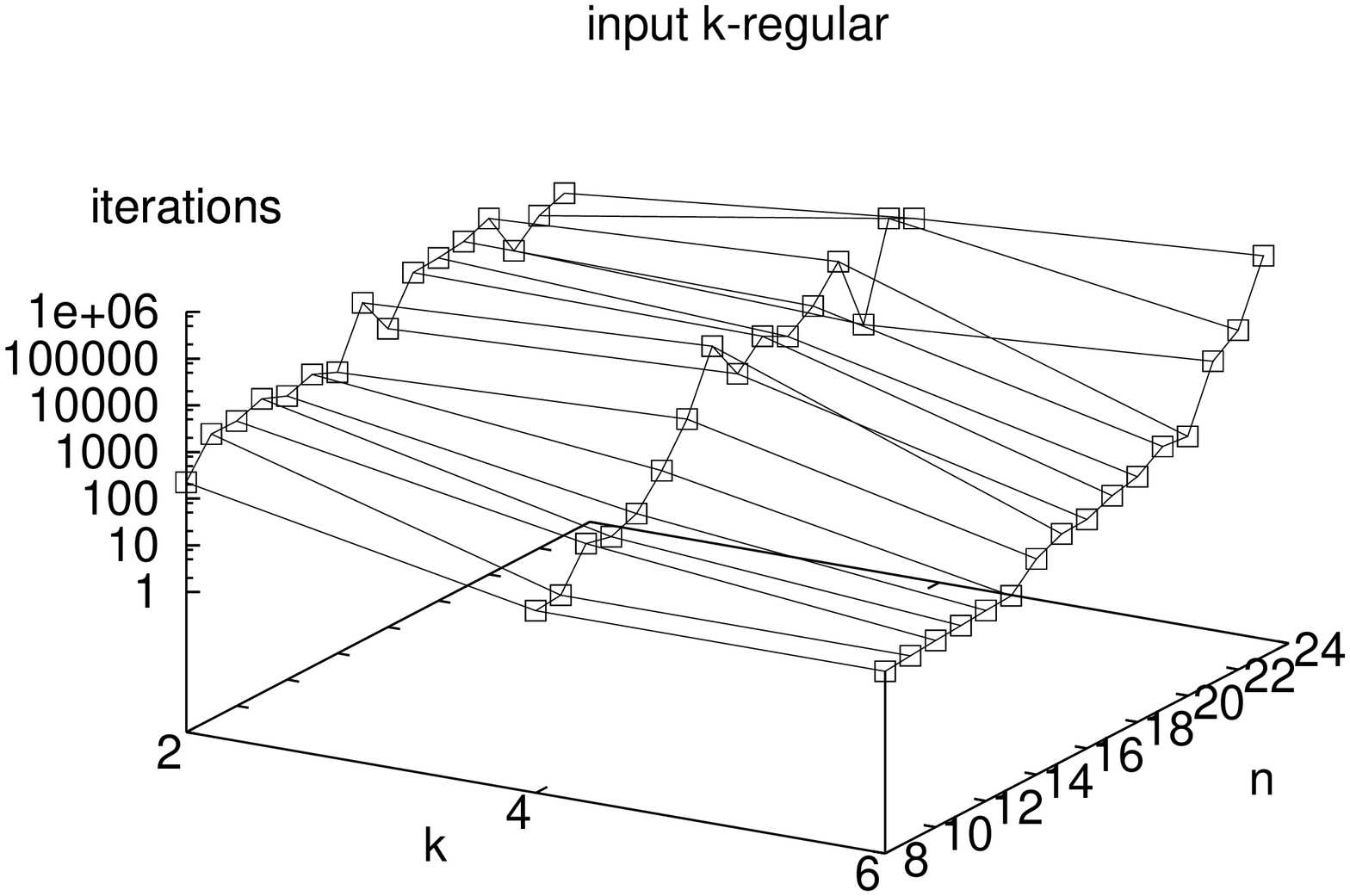}
\includegraphics[angle=-90,width=3in]{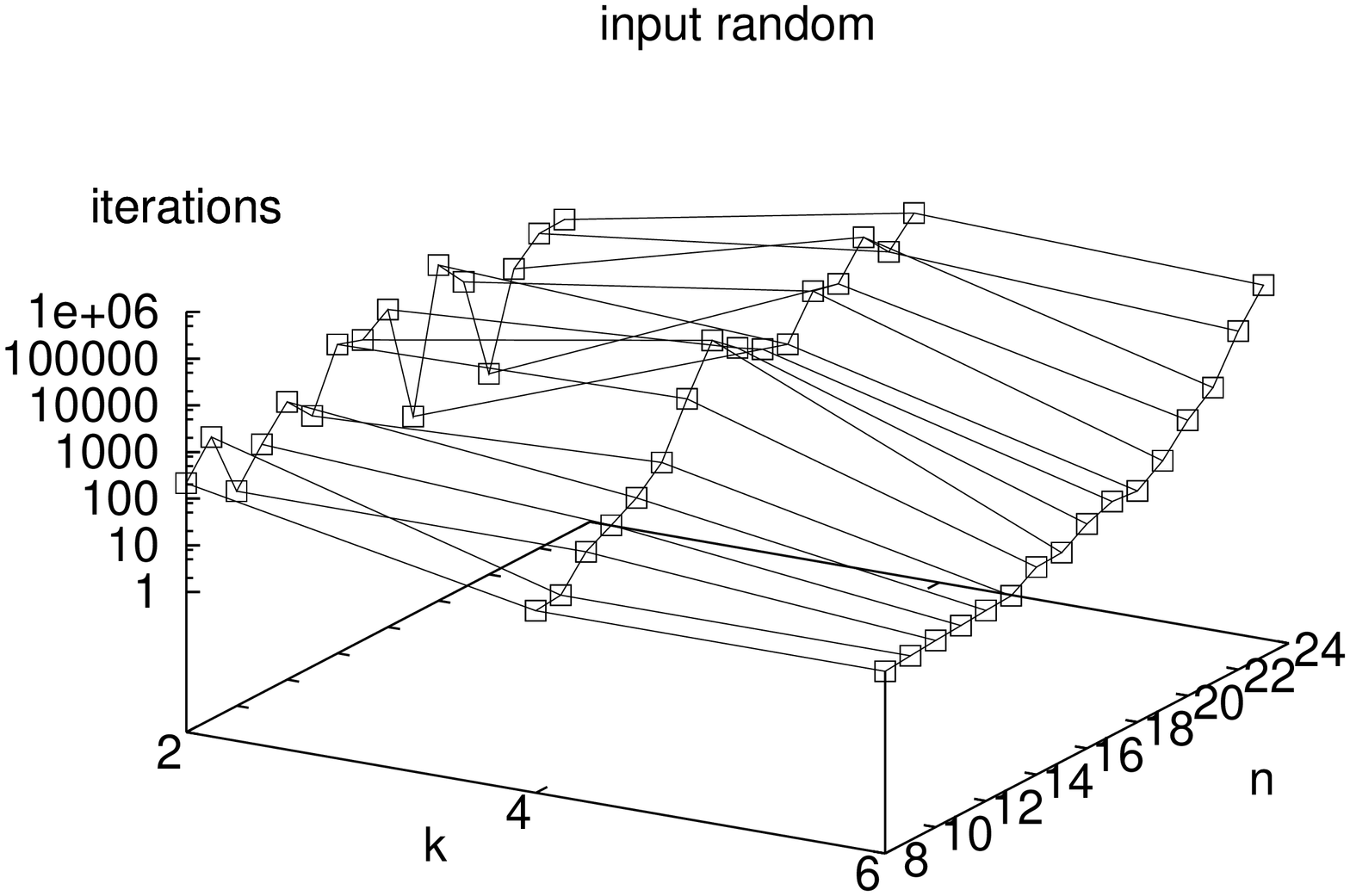}
\caption{Possible Exponential Convergence}
\end{center}
\label{fig:experimentConvergence}
\end{figure}

In the left graph of Figure \ref{fig:experimentConvergence},
  we start round-robin best-response walk from a regular $(n,k)$-wiring with
  offsets $[1:k]$.
All our experiments converge to a stable wiring.
We plot the lengths of walks for all $(n,k)$-pairs.
In the right graph of Figure \ref{fig:experimentConvergence},
  we repeat the same experiment starting from a wiring
  constructed as follows:
  Starting from a simple directed Hamiltonian cycle,
  we add to every vertex $k-1$ random out-going links.
Both experiments demonstrate lengthy and possible exponential convergence.
Moreover, the ``random'' wiring experiment shows large variance
  in the length of convergence, especially for sparse wirings.
}

\subsection*{Fairness vs Stability}
The second set of the questions concerns the graph structure of stable
uniform $(n,k)$-wirings.  Recall that a $(n,k)$-wiring is {\em
completely fair} if the costs of all vertices are the same.  The
following questions remains open:
\begin{itemize}
\item [Q5]
Is there a completely fair and stable $(n,k)$-wiring for all $n>1$ and
  $k\in [1:n]$?
\end{itemize}

We have shown that for $k > 1$, there exists an $n_{0}$ depending only
  on $k$ such that for all $n > n_{0}$, there is no stable regular
  $(n,k)$-wiring.  This result inspires us to conjecture
\begin{conjecture}\label{conj:fairstable}
For every $k>1$, there exists an $n_{0}$ depending on $k$, such that
for all $n\geq n_{0}$, there is no completely fair and $(n,k)$-stable
graph.
\end{conjecture}

A natural class of completely fair wirings is the class of
vertex-transitive graphs.  We have shown in
Section~\ref{sec:uniform.structure} that a subclass of vertex transitive
graphs, the Abelian Cayley graphs, are not stable for $k
\ge 2$.  Proving the following conjecture might be
the first step to establish~\ref{conj:fairstable}.

\begin{conjecture}\label{conj:transitive}
For every $k>1$, there exists $n_{0}$ depending on $k$, such that for
  all $n\geq n_{0}$, there is no vertex transitive directed graph over
  $[0:n-1]$ with out-degree $k$.
\end{conjecture}

Our experiments seems to suggest the following conjecture, which,
provides a potential approach to settle
Conjectures~\ref{conj:fairstable} and~\ref{conj:transitive}.

\begin{conjecture}
For every $k>1$, there exists $n_{0}$ depending on $k$, such that for
all $n\geq n_{0}$, in every pure Nash equilibrium $(n,k)$-wiring, all
but $k$ vertices have in-degree $1$.
\end{conjecture}

\junk{
\section{The Stability of Vertex Transitive Wirings}\label{sec:}

In this section, we consider the stability of a special
  class of directed graphs.
Each graph in the class has the property that
  it is vertex transitive.
This property would imply
  that the cost of all vertex are the same.
We will referred to a $(n,k)$-graph in which the
  cost of all vertex are the same as a {\em completely
  fair} $(n,k)$-graphs.

\begin{proposition}\label{pro:}
For all $n>2$ and $k\geq n/2$,
  there exists a completely fair and stable $(n,k)$-graph.
\end{proposition}
\begin{proof}
Ravi's argument.
\end{proof}

Suppose we have $n$ vertices and $k \in [1:n]$ is
  an integer.
We also assume our vertices are $V = [0:n-1]$.
A regular $(n,k)$ graph is defined by
  $k$ displacements $[b_{1},\dotsc ,b_{k}]$, where
  the edge set is
\[
E_{[b_{1},\dotsc ,b_{k}]} = \{(i,j) \ | \ (j - i \mod n) \in
\{b_{1},\dotsc ,b_{k} \} \}.
\]

We denote this directed graph by $G_{[b_{1},\dotsc ,b_{k}]} =
  (V,E_{b_{1},\dotsc ,b_{k}})$.
Clearly, a $1$-regular graph $G_{[b_{1}]}$ is either a directed cycle
when $GCD
(b_{1},n) = 1$ or a collection of directed cycles when $n$ is a
  multiple of $b_{1}$.
As the cycle is only strongly connected $(n,1)$ graph,

\begin{proposition}\label{pro:}
For all $n$ and for any $b_{1} \in [1:n-1]$, if $GCD (n,b_{1})=1$,
  then the $1$-regular graph $G_{[b_{1}]}$ is stable.
\end{proposition}

\begin{theorem}
For any $k>4$, the hypercube wiring of $(2^{k},k)$-wiring is not stable.
\end{theorem}
}

\bibliographystyle{plain}
\bibliography{RS}
\end{document}